\documentclass{aa}  

\usepackage{graphicx}
\usepackage{txfonts}
\begin{document}

   \title{ Dust attenuation  up to $z\simeq 2$ in the $AKARI$ North Ecliptic Pole Deep Field}


   \author{V. Buat
          \inst{1}
          \and N. Oi \inst{2}
            \and{S. Heinis\inst{3}}
            \and{L. Ciesla\inst{4}}
          \and D. Burgarella \inst{1}
               \and{H. Matsuhara}\inst{2}
           \and K. Malek \inst{5,6}
              \and T. Goto\inst{7}
                \and M. Malkan\inst{8}  
                \and L. Marchetti\inst{9}
                \and Y. Ohyama\inst{9}
                \and C. Pearson\inst{10,11,12}
             \and S. Serjeant\inst{11}
            }
 \institute{Aix-Marseille  Universit\'e,  CNRS, LAM (Laboratoire d'Astrophysique de Marseille) UMR7326,  13388, Marseille, France\\
              \email{veronique.buat@lam.fr}
              \and Institute of Space and Astronautical Science, Japan
Aerospace Exploration Agency, Sagamihara, Kanagawa 229-8510, Japan
 \and Department of Astronomy 
CSS Bldg., Rm. 1204, Stadium Dr. 
University of Maryland 
College Park, MD 20742-2421 
           \and University of Crete, Department of Physics and Institute of Theoretical \& Computational Physics, GR-71003 Heraklion, Greece
           \and Division of Particle and Astrophysical Science, Nagoya University, Furo-cho, Chikusa-ku, Nagoya 464-8602, Japan
           \and National Centre for Nuclear Research, ul. Hoza 69, 00-681 Warszawa, Poland
           \and Institute of Astronomy, National Tsing Hua University, No. 101,
Section 2, Kuang-Fu Road, Hsinchu, Taiwan 30013, R.O.C
  \and University of California, Los Angeles, CA 90095-1547, USA
    \and Academia Sinica, Institute of Astronomy and Astrophysics, No.1, Sec. 4, Roosevelt Rd, Taipei 10617, Taiwan, R.O.C.
  \and RAL Space, CCLRC Rutherford Appleton Laboratory, Chilton, Didcot, Oxfordshire OX11 0QX, UK
               \and The Open University, Milton Keynes, MK7 6AA, UK 
            \and Oxford Astrophysics, Denys Wilkinson Building, University of Oxford, Keble Rd, Oxford OX1 3RH, UK
           }

   \date{Received; accepted}

 
  \abstract
 {}
   {We aim to study the evolution of dust attenuation in galaxies selected in the infrared (IR) in the redshift range in which  they are known to dominate the star formation activity in the universe. The comparison with other measurements of dust attenuation in samples selected using different criteria  will give us a  global  picture of the attenuation at work in star-forming galaxies and its evolution with redshift.  }
   {We  selected galaxies in the mid-IR  from the deep survey of the North Ecliptic Field performed by the $AKARI$ satellite. Using  multiple  filters of  IRC instrument,   we  selected more than 4000 galaxies   from their rest-frame  emission at 8$\mu$m, from $z\simeq 0.2$ to $\sim 2$. We  built  spectral energy distributions from the rest-frame ultraviolet (UV)  to the   far-IR   by adding ancillary data in the optical-near IR and from $GALEX$ and $Herschel$ surveys.  We fit  spectral energy distributions  with  the physically-motivated code CIGALE. We test  different templates for active galactic nuclei (AGNs) and recipes  for dust attenuation and  estimate stellar masses, star formation rates,  amount of dust attenuation, and  AGN contribution to the  total IR luminosity. We discuss the uncertainties affecting these estimates on a subsample of galaxies with spectroscopic redshifts. We also define a subsample of galaxies with  an IR luminosity close to the characteristic IR luminosity  at each redshift and study  the evolution of dust attenuation of this selection representative of the bulk of the IR emission.}
   {The AGN contribution to the total IR luminosity   is found to be on average approximately 10$\%,$   with a slight increase with redshift. The determination of  AGN\ contribution does not depend significantly on  the assumed AGN templates  except for galaxies detected in X-ray. The choice  of  attenuation law has a marginal impact on the determination of stellar masses and star formation rates.   Dust attenuation in galaxies dominating the IR luminosity function  is found to increase from $z=0$ to $z=1$ and to remain almost constant from $z=1$ to $z=1.5$. Conversely, when galaxies are  selected at a fixed  IR luminosity,  their dust attenuation slightly  decreases as redshift increases but with a large dispersion, confirming previous results obtained at lower redshift.The attenuation in our  mid-IR selected sample  is found $\sim 2$ mag higher than that found  globally in the universe or in UV and H$\alpha$ line selections in the same redshift range.   This difference is well explained by an increase  of dust attenuation  with the stellar mass, in global agreement with other recent studies.  Starbursting galaxies do not systematically exhibit a high attenuation. We conclude that the galaxies selected in IR and dominating the star formation exhibit a higher attenuation than those measured on average in the universe because  they are massive systems. Conversely UV selected galaxies exhibit a large range of stellar masses leading in a lower average attenuation than that found in an IR selection.}
   {}


   \maketitle
%

\section{Introduction}
One of the most complex  processes in galaxies is star formation and understanding it and its evolution with time remains a challenge for modern astronomy. The evolution of dark matter is now modeled with  high accuracy and is shaping the large scale structures in the universe. The physics of baryons, including star formation,  is much more difficult to understand and  to implement in models of galaxy formation and evolution. The first natural step is to find constrains from observations by measuring a  reliable star formation rate (SFR)  at various scales and redshifts and studying the main drivers of its variations.

Major progress has been made in the measure of the SFR inside the disk of nearby galaxies with   strong relation found between SFR and molecular gas content \cite[e.g.,][for a review]{bigiel08,leroy08,kennicutt12},   and the measure of the star formation efficiency in galaxies is now possible up to large redshifts \cite[e.g.,][]{daddi10,tacconi10,tacconi13,santini14}. Besides these studies aimed to understand   the process of star formation, the global amount of star formation in the universe is measured by building statistical samples with observables related to the recent star formation. The rest-frame UV  emission is frequently privileged. The observations of the $GALEX$ satellite up to $z=1$ and the numerous optical surveys give very sensitive UV rest-frame measurements from the nearby universe  to high redshifts and for large samples of galaxies \citep[e.g.,][]{salim06,cucciati12,finkelstein12}. However , UV emission suffers from a major issue, which is dust attenuation. In the nearby universe and  within galaxies  like the Milky Way, about half of the stellar light is absorbed and re-emitted by dust at wavelengths larger than $\rm \sim 5 \mu m$, and that fraction can increase to more than 90$\%$ in  starbursting objects. Therefore  measuring the IR emission of galaxies has been identified as mandatory for measuring the total SFR, and adding both UV and IR  emissions is now recognized as a very robust  method to measure the SFR. The development of IR facilities allows the building of statistical samples of galaxies detected in IR, although the lower sensitivity of IR detectors combined with poor spatial resolution makes comparison with UV-optical surveys  still difficult.  The $\it Spitzer$ and $\it Herschel$ deep observations combined with ground-based optical surveys give us a complete view of the star formation  up to $z\sim 2$  \citep[e.g.,][]{reddy08,buat12,reddy12,oteo14} although selection biases are recognized to be  complex \citep{bernhard14,heinis14}. The very high redshifts ($z>>2$) remain almost unexplored in IR even with $Herschel$ \cite{madau14}  and  stacking methods are intensively used to find the average IR  emission of optically faint objects not detected individually \citep[e.g.,][]{reddy12,ibar13,heinis13,heinis14,pannella14,whitaker14}.\\
To complement  the analysis of galaxy samples, the comparison of  UV and IR luminosity functions gives the global energy budget of the universe.  \citet{takeuchi05} first compare the UV and IR luminosity functions from $z=0$ to $z=1,$ and \citet{burgarella13} extend the analysis up to $z=4$ taking advantage of the most recent $Herschel$ measurements. Both studies  measure the  evolution with redshift of the global dust attenuation  by comparing the UV and IR luminosity densities:  they found it increases  from $z=0$ to $z=1.2$ and then  declining up to z=4 with a similar amount of dust attenuation at $z=0$ and $z=4$.   Although these results depend on the exact shape of the IR luminosity function, which remains uncertain especially at high redshift \citep[][for a review]{casey14}, they   are found to be very consistent with those obtained by  \citet{cucciati12} for a large sample  of UV selected galaxies  in the same redshift range. \\
UV and IR luminosity functions exhibit very different shapes \citep[][and references therein]{burgarella13}: bright objects are much more numerous in IR than in UV and the luminosity evolution with redshift is stronger in IR than in UV. The faint end of the IR luminosity function is not well constrained but may be considerably flatter than that measured for the UV luminosity function \citep{gruppioni13}. These very different distributions imply that UV or IR selected samples are expected to be composed of different galaxy populations, in terms of luminosity and dust attenuation \citep{buat06}.  \\
The strong impact of dust attenuation and the need to estimate it as accurately as possible when UV data are used has led to numerous investigations of   samples of star-forming galaxies. Quite naturally these  studies focus on  galaxies selected either in their UV rest-frame or from their emission in gas recombination lines. A correlation is found between the amount of dust attenuation and the stellar mass of these galaxies at low and high redshift \citep{garn10,sawicki12,buat12,ibar13,kashino13}. \citet{heinis14} study the IR properties of UV selected galaxies in the COSMOS field  by stacking Herschel/SPIRE images at redshift 1.5, 3 and 4. They find that the average amount of dust attenuation    correlates well with the stellar mass in their sample whereas it remains roughly constant when  the observed UV luminosity ($L_{\rm UV}$) varies.  For a given  $L_{\rm UV}$ , the galaxies  exhibit a very large range of masses and thus  dust attenuation.\\

The aim of the present work is to select galaxies in the IR to investigate the impact of this kind of  selection on  the dust attenuation properties and to compare our results  to the ones  obtained from  UV rest-frame selections or globally in the universe. It is of particular importance since most of the star formation in the universe can be securely measured in IR up to redshift $\sim 2$ since the stellar  emission reprocessed by dust dominates the unattenuated emission from stars. Numerous studies were already devoted  to characterize dust attenuation in IR selected galaxies \citep[e.g.,][]{goldader02, burgarella05, buat05, howell10, takeuchi10}. A major result of these studies is the deviation of the sources detected in IR surveys  from the relation found between the ratio of IR to UV luminosity and the slope of the UV continuum  for local starburst galaxies  with bluer colors than expected  (but see also \citet{murphy11}). Recently \citet{casey14b} attributed this departure to a patchy geometry. \\
The infrared space telescope $AKARI$ carried out a deep survey of the North Ecliptic Pole (hereafter NEP-deep) with all the filters of the InfraRedCamera (IRC). We take advantage of the continuous filter coverage in the mid-IR   to build a 8$\mu$m rest-frame selection following the strategy of \citet{goto10}. A 8 $\mu$m selection is  relevant to select  dusty galaxies  active in star formation since it focuses on  the emission of polycyclic aromatic  hydrocarbons. Main sequence galaxies defined to have a normal mode of star formation and a tight relation between their star formation rate and their stellar mass are found to have similar ratios of 8$\mu$m rest-frame  to total IR luminosities \citep{elbaz11,nordon12,murata14}. For these galaxies, a 8$\mu$m selection is similar to a selection based on their total IR emission. Using the S11, L15, L18W, and L24 filters, we can select galaxies properly on a large range of redshift  from $z=0.15$ to $z=2.05$. Combining the near to mid-IR catalog of \citet{murata13} with the optical data of \citet{oi14} and with $GALEX$ and $Herschel$/PACS detections, we are able to build the UV to IR spectral energy distributions (SED) of our selected sources. These SEDs are analyzed with the   code CIGALE  to measure physical parameters such as dust attenuation, SFR, or stellar masses.  We first optimize the SED fitting process on a subsample of galaxies with a full wavelength coverage and spectroscopic redshifts and then we run CIGALE on the whole sample of 4077 galaxies. The amount of dust attenuation and its evolution in redshift is measured  and compared to the results found with other selections.\\

The paper is organized as follows: the sample selection is described in Sect. 2, and the SED fitting process and its optimization on a subsample of galaxies is detailed in Sect. 3. The main result of this work, the measure of dust attenuation and its evolution with redshift, is presented in Sect. 4. In Sect. 5, we compare our results to other measures of dust attenuation for different sample selections  and propose the stellar mass as the main driver for dust attenuation. Our conclusions are presented in Sect. 6.\\
Throughout the paper we use the WMAP7 cosmological parameters. The parameter $L_{\rm IR}$  corresponds to the total  luminosity  emitted by dust as defined in Sect. 3, and the UV luminosity $L_{\rm UV}$ is defined as $\nu L_{\nu}$ at 150 nm. All luminosities are expressed in solar units ($L_{\odot} = 3.83 10^{33} \rm {erg s^{-1}}$).
\begin{table*}
\caption{Sample selection. Central wavelength and FWHM of the filters are expressed in $\mu$m. The last column (UV data) is the number of galaxies for which at least one measurement in the UV continuum ($< \simeq  0.25 \mu$m) is available. The filters corresponding to the UV range are indicated in brackets.}
\label{tab:sample}
\begin{tabular}{c c c c c c c c}
\hline
Bin &Filters &  $\lambda_{cent}$ & FWHM & Redshift range& Number of galaxies & PACS matches& UV data\\
\hline\hline
 1 & S11&10.61&4.08& $0.15<z<0.49$&1661&352& 1031 (NUV)\\
 2 &L15 & 15.98 & 6.5 & $0.75<z<1.34$&1906&190&1748 (u,g)\\
 3 & L18  & 19.6 & 11.2 & $1.34<z<1.85$&460&50&409 (u,g,r)\\
4 & L24  & 23.11 & 5.25 & $1.85<z<2.05$&50&7&45 (g,r)\\
\hline
\end{tabular}
\end{table*}
\section{Sample selection}

\begin{figure}
   \centering
  \includegraphics[width=\columnwidth]{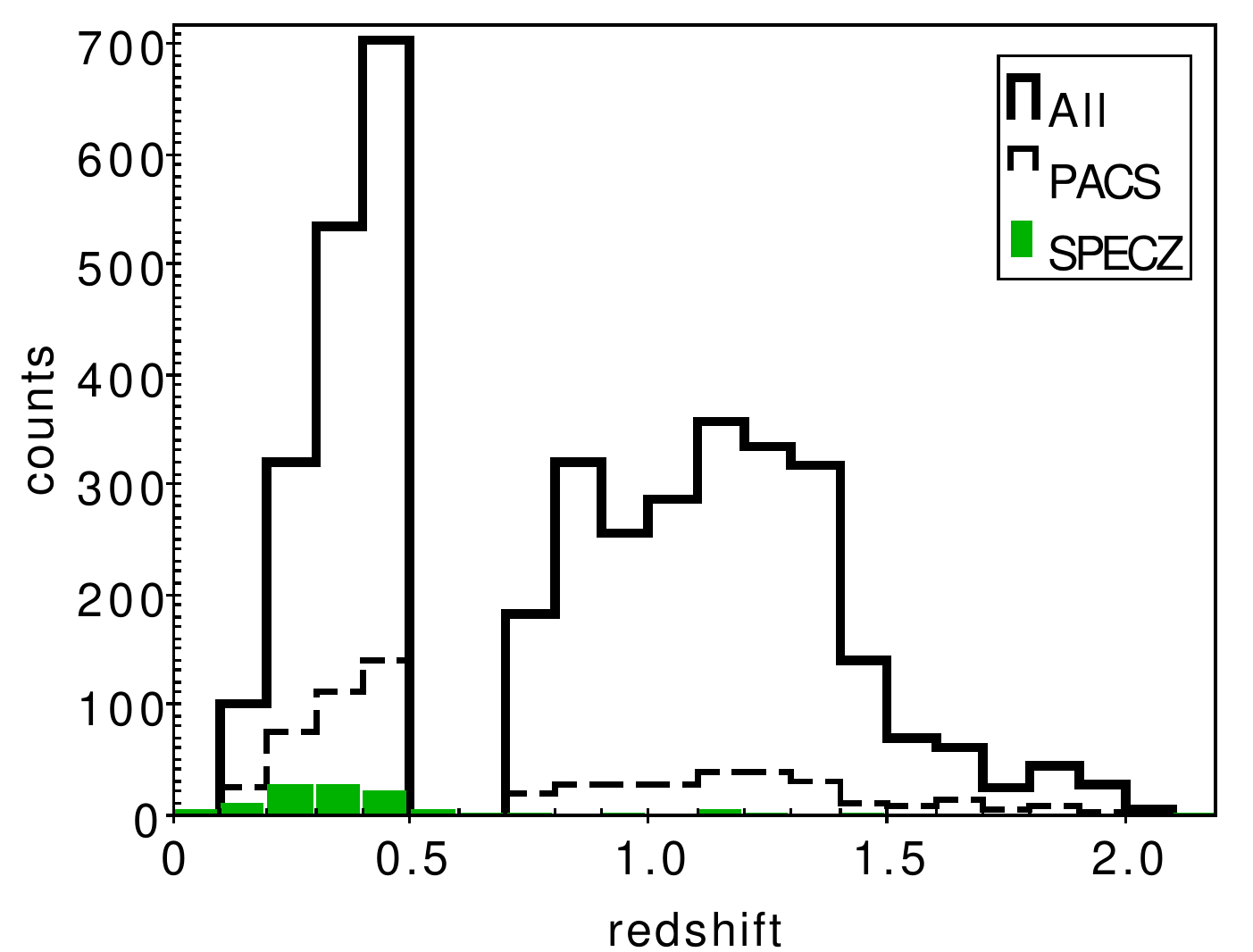}
   \caption{Redshift distribution of the whole sample (black solid  line) of galaxies detected with PACS (black dotted line) and with a spectroscopic redshift (SPECZ sample: green-filled histogram). }
              \label{zdist}%
    \end{figure}

Our aim is to combine data at different wavelengths for galaxies primarily selected at 8$\mu$m in their rest frame.
We start with the \citet{murata13} catalog of the $AKARI$-NEP deep survey, which   contains 27770 sources over 0.5 deg$^2$. We cross-matched this catalog with the optical-near-IR catalog of \citet{oi14} for which photometric redshift are calculated. We found 23345 sources  in common within a tolerance radius of 1 arcsec on the coordinates.   MegaCam ($\rm u^{\star},g',r',i',z'$) and WIRCam (Y,J,Ks) data are available for these sources. The 5$\sigma$  detection limits correspond to $\rm u^{\star} = 24.6, g' = 26.5, r' = 25.7, i' = 24.9, z' = 23.9, Y = 23.2, J = 22.8,$ and $\rm Ks = 22.5$ mag (AB scale). The photometric redshifts are estimated using {\sl Le Phare} software \citep{ilbert06}. The  redshift accuracy ($\sigma_{\Delta z/(1+z)}$) is found to be equal to 0.032 at $z<1$ and $0.117$ at $z>1$. We refer to \citet{oi14} for a more  detailed description. \\
We perform our 8$\mu$m rest-frame selection following the same strategy as \cite{goto10}. We consider the four $AKARI$ filters S11, L15, L18 and L24 and define redshift bins so that the observed wavelength of the 8$\mu$m feature corresponds to a transmission of the filters larger than 0.8.  We then check that the detection rate  in the photometric catalog of \citet{oi14} is similar for each subsample of sources detected in one of these these filters  (85$\%$ for sources detected in  the S11 filter, 80$\%$ for the L15 and L18 filters, and   78$\%$ for the L24 filter), which ensures us that we do not have any systematic bias between our bins. The L18 filter has a very broad bandpass, which largely overlaps those of the L15 and L24 filters. We  first define the redshift range reachable with the L15 filter, and then the redshift bin of the L18 filter starts at $z = 1.34$  in order not to overlap the one defined with the L15 filter. The L24 band has a low sensitivity and we prefer to use it only when  the 8$\mu$m rest-frame comes out of the L18 filter ($z>1.85$). We define a sample of 4077 galaxies. The  redshift distribution is described in Table 1 together with the number of galaxies per bin of redshift defined with each filter.\\
The $AKARI$-NEP  field was also observed by $GALEX$ for 89450 s (Program GI4-057001-AKARI-NEP, P.I. M. Malkan). When available we add the $GALEX$/NUV data for galaxies with redshift lower than 0.925, corresponding to a rest-frame wavelength larger than 1200 $\AA$. The NUV fluxes were measured from the images taken on the MAST database, the flux extraction is  performed with DAOPHOT   (Mazyed et al., in preparation) and the detected sources are cross-matched with the optical coordinates of the sources within a matching radius of 2 arcsec. NUV fluxes are added to the SED of 1207 sources with $z<0.925$.\\
The $AKARI$-NEP field was also observed by $Herschel$ with the PACS and SPIRE instruments  for 73.5 hours (Program OT1-sserj01-1, P.I. S. Serjeant). Given the small number of sources detected with SPIRE in the  field studied in this work,  we only consider PACS data.  Fluxes  at 100 and 160 $\mu$m  from the PACS images were also measured with DAOPHOT (Mazyed et al., in preparation) the  match with our selected sources catalog is made with a matching radius of 3.5 arcsec, we keep only sources with a single match within this radius. We find that 599 sources are detected with PACS at 100 $\mu$m and 270, both at 100 and 160 $\mu$m (Table 1). \\
We are interested in studying  dust attenuation   in these galaxies. To this aim, data in the UV range bring information very complementary to those in the IR range. Therefore we check the number of sources for which rest-frame UV data are available.  In Table 1  we  report  the number of sources in each redshift bin with a least one detection at a wavelength lower than 0.25 $\mu$m in the rest frame of the galaxy. Approximately 90$\%$ of the sources are observed in UV except for the first redshift bin for which  only 62$\%$ of the selected sources are detected with $GALEX$.\\
  The $AKARI$-NEP  Deep  Field was observed with $Chandra$ with a detection limit of $\rm \sim 10^{-15} erg s^{-1} cm^{-2}$ in the 0.5-2 keV band. The X-ray source catalog is presented in \citet{krumpe14}. We used the V1.0 version of the catalog. Unfortunately  X-ray observations cover only approximately half of our field. We    cross-matched the catalog of X-ray sources  from \cite{krumpe14} with our selection following their prescription (separation lower than the match radius) and found 147 matches. It corresponds to 6$\%$ of our initial sample. We can expect the same low contribution of such  X-ray sources for the whole field. \\
 Last,  we also define a subsample of our 8$\mu$m selected galaxies with a spectroscopic redshift  from the compilation of \citet{oi14}  and detected with PACS as well as in most of the other bands: 106 galaxies are thus selected, 7 are classified AGN type 1, and 1 is classified AGN  type 2. The field where spectroscopic redshifts are available is entirely covered by the X-ray observations. Fifteen sources are  detected, and four   out of the seven AGN type 1 and  AGN type 2 are also detected in X-ray. This catalog is called  SPECZ sample in the following and is used to optimize the SED fitting process. The redshift distribution of the sources included in each sample is shown in Fig.\ref{zdist}.

\section{Fitting the spectral energy distributions}

The SED fitting is performed with the  version v0.3 of the CIGALE code (Code Investigating GALaxy Emission)\footnote{http://cigale.lam.fr} developed with PYTHON.  CIGALE combines a UV-optical stellar SED with a dust component emitting in the IR and fully conserves the energy balance between  dust absorbed  emission and its re-emission in the IR.  This energy balance is also observed   for  the AGN emission absorbed and re-emitted by the dusty torus (see below). The total IR luminosity $L_{\rm IR}$ is defined as the sum of  luminosities coming  from stellar and AGN light re-processed by dust.  Star formation history as well as  dust attenuation characteristics, including the attenuation law, are input parameters that can be either taken free or fixed according to the available data or the specific aims.  The main characteristics of the code are described in \cite{noll09}. We refer to Burgarella et al. (in preparation) and Boquien et al. (in preparation) for a detailed description of the new version  of the code. Here we only describe  the assumptions and choices specific to the current study. The main parameters and  range of  input values are reported in Table 2. The output values parameters are estimated by building the probability distribution function (PDF)  and by taking the mean and standard deviation of the PDF.
\begin{table}
\caption{Range of values of the  input parameters  for dust and stellar emission, used for the SED fitting with CIGALE.}
\label{tab:parameters}
\centering
\begin{tabular}{l  l}
\hline
Parameter & Range \\
\hline\hline
Amount of dust attenuation $E(B-V)$ \tablefootmark{1}   & 0.1-1\,mag \\
Attenuation curve & B12,C00, SMC-like\\
IR templates\tablefootmark{2}, $\alpha$& 1-3\\
AGN fraction,$\rm frac_{AGN}$& 0-0.5\\
\hline
Stellar populations  &  \\
\hline
Age (old stellar population)  $t_f$ \tablefootmark{3}&  2-11\,Gyr \\
$e$-folding rate (old stellar population)  $\tau$ &  1-5\,Gyr \\
Age  (young stellar population) $t_{\rm ySP}$ &  50-500\,Myr \\
Stellar mass fraction \tablefootmark{4}  $f_{\rm ySP}$ &  0.01-0.2\\
\hline
\end{tabular}
\tablefoot{\\
\tablefootmark{1}{$E(B-V)$ corresponds to the attenuation of the youngest population and   a reduction factor of 0.5 is applied to the color excess of stellar populations older than $10^7$ years.}\\
\tablefootmark{2} {$\alpha$ is the exponent of the power-law distribution of dust mass over heating intensity \citep{dale14}.}\\
  \tablefootmark{3}{$t_f$ is always lower than the age of the universe at the redshift of the source.}\\
  \tablefootmark{4}{Stellar mass fraction produced with the young stellar population.}
  }
\end{table}

\subsection{Star formation history}
Different scenarios of star formation history are implemented in CIGALE: exponentially declining or rising SFR, delayed SFR, and a declining SFR with an overimposed burst. It is possible to study any SFR by adding a file. As shown in several studies \citep[e.g.,][and references therein]{pforr12,buat14,conroy13}, it is difficult to disentangle the different scenarios  because of  the degeneracy of the fits. As long as IR data are available and  we are dealing with galaxies that are actively forming stars,  the measure of dust attenuation is not affected by the detailed star formation history \citep{meurer99,buat05,cortese08,buat11,hao11}.   We adopt a model with  two stellar populations:  a recent stellar population with a constant SFR on top of an older stellar population created with an exponentially declining SFR. Ages of the older stellar population and young component ($t_f$  and  $t_{\rm ySP}$ respectively)  are free parameters  \citep{noll09}. The initial mass function of \cite{chabrier03} is adopted  with  the stellar synthesis models of  \citet{bc03}. Compared to the widely used, simple exponentially decreasing SFR, this type of bimodal star formation history is able to better reproduce real systems with several phases of star formation known and to give more realistic stellar ages  \citep[][and references therein]{buat14}.

\subsection{Dust attenuation recipe and dust re-emission}
Since we are fitting the SED of galaxies selected in the IR, the treatment of dust attenuation and re-emission is critical. In CIGALE, dust attenuation is controled by the choice of the attenuation curve. The light re-processed by dust and re-emitted in IR is modeled  using IR SED templates.

\subsubsection{Attenuation curves}
 To model the attenuation by dust, the code uses  the attenuation law of \citet{calzetti00} (hereafter C00), and offers the possibility of varying the steepness of this law and adding a bump centered at 2175\,$\AA$. We refer to \citet{noll09} for a complete description of the dust attenuation prescription, which still holds in the new version of the code. In brief, the dust attenuation is described as
 \begin{equation}\label{eq:attlaw}
A_\lambda  = {E(B-V)}~ (k'(\lambda) + D_{\lambda_0,\gamma,E_ {\rm b}}(\lambda)) \left(\lambda \over {\lambda_V} \right)^{\delta},
\end{equation}
where $\lambda_V = 5500\,\AA$, $k'(\lambda)$ comes from \citet[][Eq.4]{calzetti00} and $D_{\lambda_0,\gamma,E_{\rm b}}(\lambda)$, the Lorentzian-like Drude profile commonly used to describe the UV bump \citep{fitz90}, is defined as
\begin{equation}\label{eq:bump}
D_{\lambda_0,\gamma,E_{\rm b}} = {{E_{\rm b} \lambda^2 \gamma^2} \over {(\lambda^2-\lambda_0^2)^2 + \lambda^2 \gamma^2}}.
\end{equation}
The coarse wavelength coverage in the UV rest frame does not allow us to measure the dust attenuation curve for every galaxy. Instead we can run CIGALE with different predefined attenuation curves and compare the results. 
We obtained our fiducial attenuation curve by  studying high redshift galaxies  (\cite{buat12}, hereafter B12) , corresponding to $E_{\rm b} = 1.6$ and $\delta=-0.27$, which  is close to the LMC2 extinction curve with a UV bump of moderate amplitude. We also consider the C00 attenuation law  ($E_{\rm b}=0, \delta=0$).  \cite{ilbert09} found that considering an SMC extinction curve and a screen geometry could help to measure photometric redshifts. The  curve corresponding to $\delta=-0.5$ and $E_{\rm b} = 0$ is similar to the SMC curve of \cite{gordon03} \citep{boquien12} and is added for comparison (hereafter  SMC-like).
We introduce a differential attenuation for young and old stars, the color excess is reduced by a factor 2 for stars older than $10^7$ years \citep{calzetti97,charlot00,panuzzo07}.

 \subsubsection{Dust emission templates}
The  stellar emission absorbed by dust is re-emitted  in IR. This emission is generated using different empirical templates. \citet{draine07}, \citet{casey12}, and  \citet{dale14} models  are  implemented in CIGALE. 
The templates used in this work  are those of  \citet{dale14}. They are updated templates from the  \cite{dale02} library, which  provides  good fits of $Herschel$ data \citep[e.g.,][]{wuyts11,  magnelli13}. \cite{dale14} also add a quasar component. Here we use these templates without this quasar contribution  since the AGN contribution is  defined separately in CIGALE, as described below.
  \begin{figure}
   \centering
  \includegraphics[width=\columnwidth]{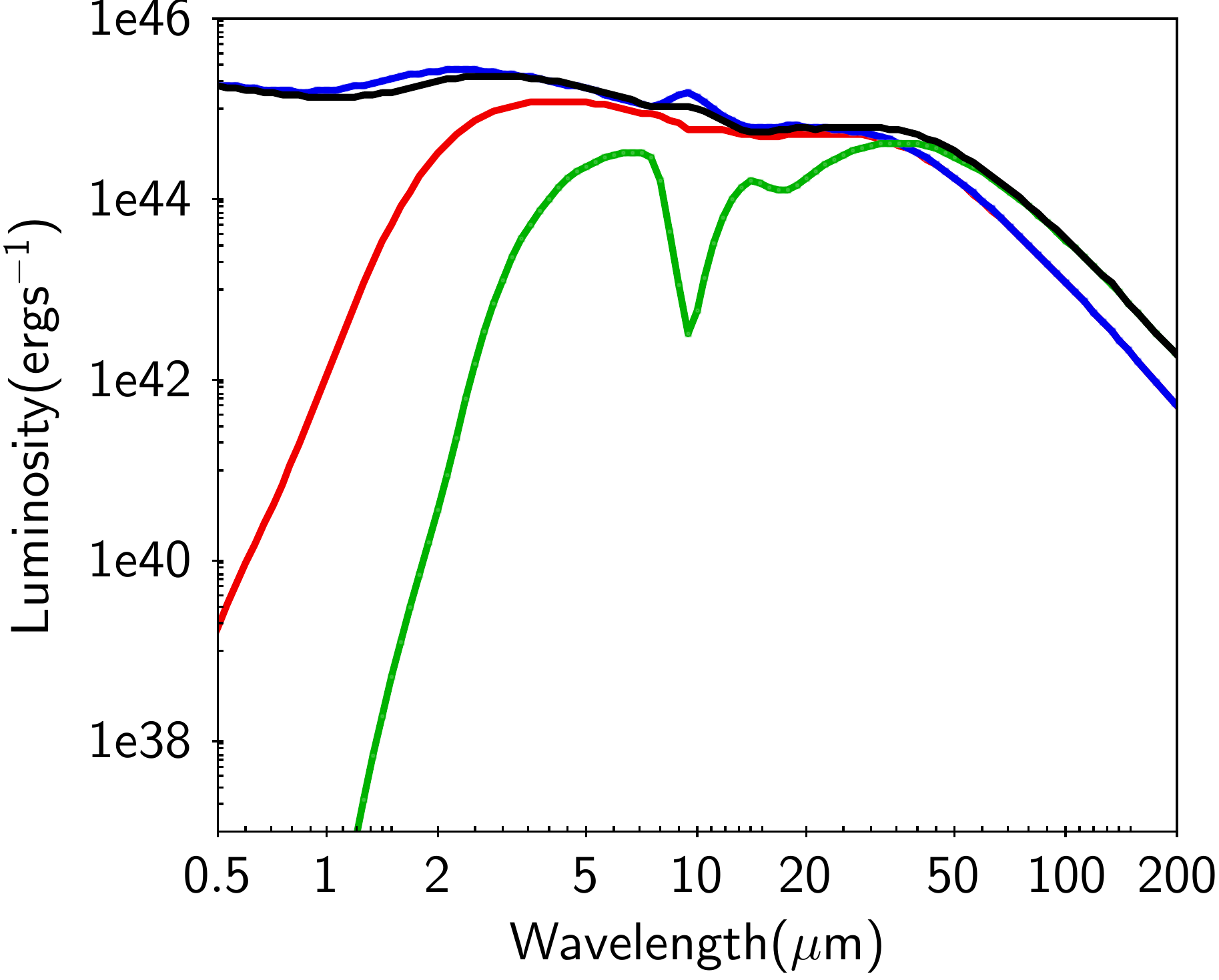}
   \caption{SED templates for AGNs considered in this work and  defined from the \citet{fritz06} library with different values of the optical depths $\tau_{9.7}$ and  viewing angle $\psi$. The other parameters are fixed as described in the text. Green line: $\tau_{9.7} = 6$, $\psi=0$; red line: $\tau_{9.7} = 1$, $\psi=0$; blue line: $\tau_{9.7} = 1$, $\psi=80$; black line: $\tau_{9.7} = 6$, $\psi=80$. The fiducial models correspond to $\tau_{9.7} = 1,6$ and  $\psi=0$. The models represented here correspond to a bolometric luminosity of $\rm 10^{46} erg s^{-1}$.}
              \label{agn}%
    \end{figure}

  \begin{figure*}
   \centering
 \includegraphics[width=18cm]{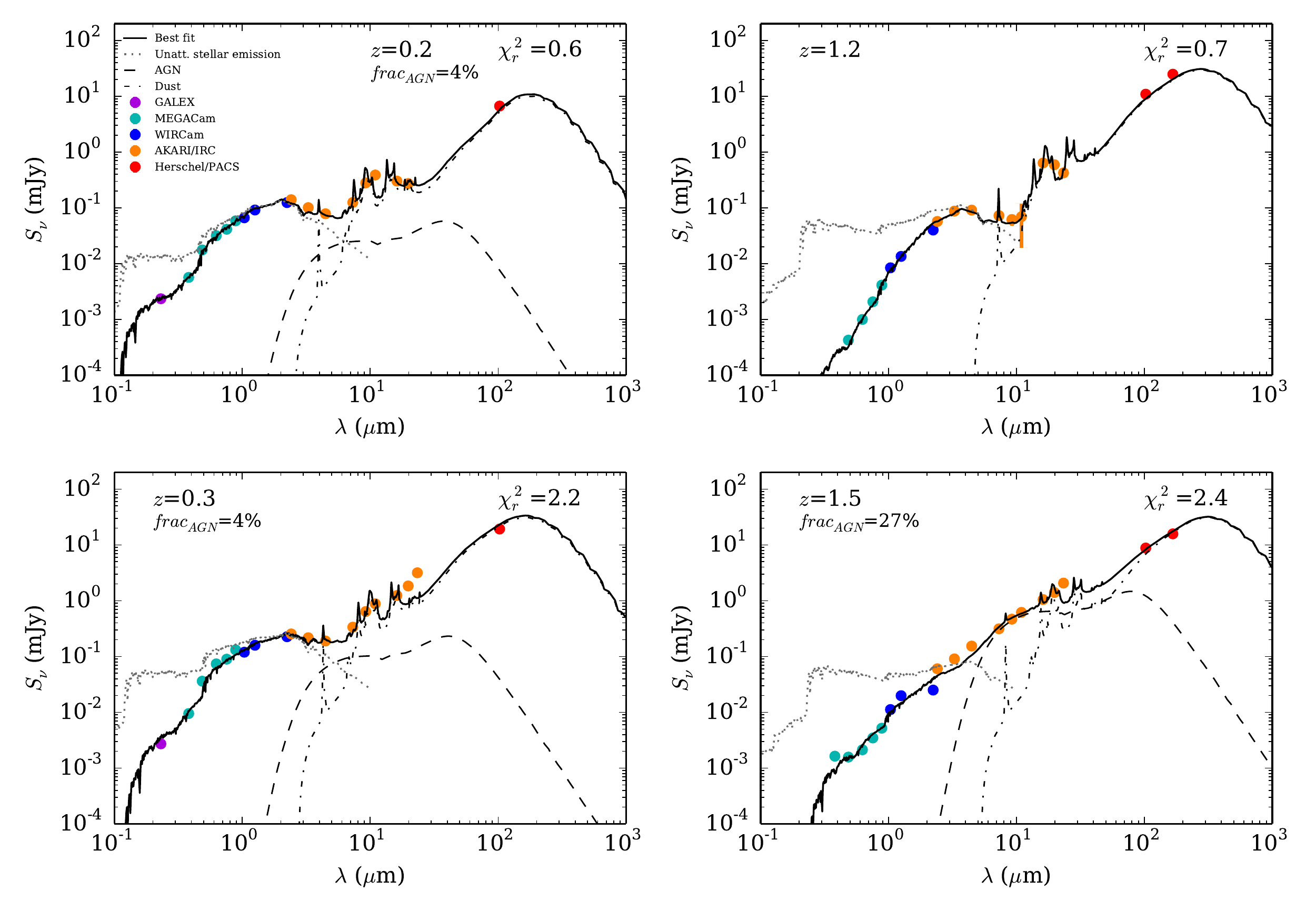}
   \caption{Examples of fits for the SPECZ sample with various redshifts, $\chi_r^2$ values, and AGN contributions. The observed wavelength is plotted on the x-axis. On the y-axis the observed fluxes are plotted with colored points.  The best model and its different components (spectra of the unattenuated stellar, dust (thermal) and  AGN emission) are plotted as lines. The different colors and lines  are described in the legend  inside the top left plot. In the top right plot the  value of $\rm frac_{AGN}$  is not reported since  we do not find any AGN component for this source.}
              \label{sed}%
    \end{figure*}
\subsection{AGN component}

With CIGALE, it is possible to add the contribution of an AGN to the SED. The adopted templates are those of  \cite{fritz06} based on two components: the emission of the central source and the radiation from the dusty torus in the vicinity heated by the central   source. \cite{fritz06}   introduce a set of six parameters  in their model, describing the geometrical configuration of the torus and  the properties of the dust and solving the radiation transfer equation. The dust distribution inside the torus is assumed to be smooth.  The different parameters are described in  \citet{hatziminaoglou08}. When one accounts for  this component in  CIGALE, all the free parameters  related to this AGN contribution are added to the already numerous parameters related to  stellar (direct and re-processed) emission. With only photometric data we are confronted by the well-known degeneracy of SED fitting processes     induced by the limited number of data points compared to  the number of  parameters used to build the SED templates. From simulated composite SEDs  with an AGN and a stellar component, \citet{ciesla14} find that only extreme values of the  angle between the AGN axis and the line of sight, characterizing the nature of the AGN (type 1 or 2), can be estimated. Thus we choose to fix several parameters to average values: from \citet{hatziminaoglou08} the ratio of the outer to inner radius is fixed to  60, the dust density parameters $\beta$  and $\gamma$ are fixed to -0.5 and 0, respectively, and the opening angle of the torus is equal to 100$\rm ^o$. The remaining free parameters are the optical depth at 9.7 $\mu$m ($\tau_{9.7}$ ) and the angle $\psi$ between the AGN axis and the line of sight. \\
  
 CIGALE calculates the relative contribution of the dusty torus of the AGN to the total IR luminosity that is called the AGN fraction ($\rm frac_{AGN}$). We explore several combinations of the three  parameters: $\tau_{9.7}$ , $\psi$ , and  $\rm frac_{AGN}$. A large number  of AGN fractions is introduced since it is the main parameter relevant to the AGN component that we want to retrieve: we consider eight values of $\rm frac_{AGN}$ linearly spaced between 0 and 0.5 (cf. Table 2). Following the prescription of \citet{ciesla14}, we consider  only two extreme 
 values for the angle $\psi$, $\rm \psi=0^o$ and 80$\rm ^o$ to reproduce an AGN type 2 and type 1, respectively\footnote{The definition  of  $\psi$  used  in the grid of  AGN models  implemented in the current version of CIGALE (V0.3) is different from the one used in the paper of \cite{fritz06}. If we note $\psi_0$ the value used by \cite{fritz06} then $\psi = 90-\psi_0$. For the sake of consistency, we use the CIGALE convention.} . After several trade-offs we also consider a low  ($\tau_{9.7}$= 1) and a high  ($\tau_{9.7}$= 6) optical depth model. \cite{hatziminaoglou08} have also introduced models  corresponding to  low optical depth tori  to improve the fits of their SWIRE/SDSS quasars. Our four selected  templates are plotted in Fig.\ref{agn}. We can see that $\tau_{9.7}$= 6 and $\rm \psi=0^o$ corresponds to a hidden AGN with a strong silicate absorption.    The two templates  corresponding to $\rm \psi=80^o$ are very similar and correspond to a quasar emission with a strong UV/optical emission. Combining this kind of an SED with a stellar component, even with a low AGN fraction, will induce a high contribution of the AGN.
 The template corresponding to $\tau_{9.7}$= 1 and $\rm \psi=0^o$ exhibits a smooth distribution in the mid-IR, not very different from the cases with  $\rm \psi= 80^o$ , if we ignore the small silicate emission of these templates, which does not affect broadband measurements. However, in this case, the UV-optical part is obscured since  crossing the torus leads to a much lower flux than in the case $\rm \psi=80^o$.  The AGN models with a smooth dust distribution are known to produce sudden jumps in the SEDs, and whether the line of sight intercepts the torus or not, a clumpy dust distribution will produce a smoother transition \citep{feltre12}.  So as not to  introduce an overly large contribution of the AGN in the UV-optical, but to allow for a silicate absorption or a smoother distribution in the mid-IR, our  fits will be based  on two configurations: $\tau_{9.7}=1,6$ and $\rm \psi=0^o$ for  scenario 1,   $\tau_{9.7}=6$ and $\rm \psi=0,80^o$ for  scenario 2\footnote {We did not consider the four AGN templates in a same run to avoid putting too large weight to  $\psi=80^o$, the two SEDs corresponding to this value of $\psi$ being very similar.}.

\begin{figure}
   \centering
  \includegraphics[width=9cm]{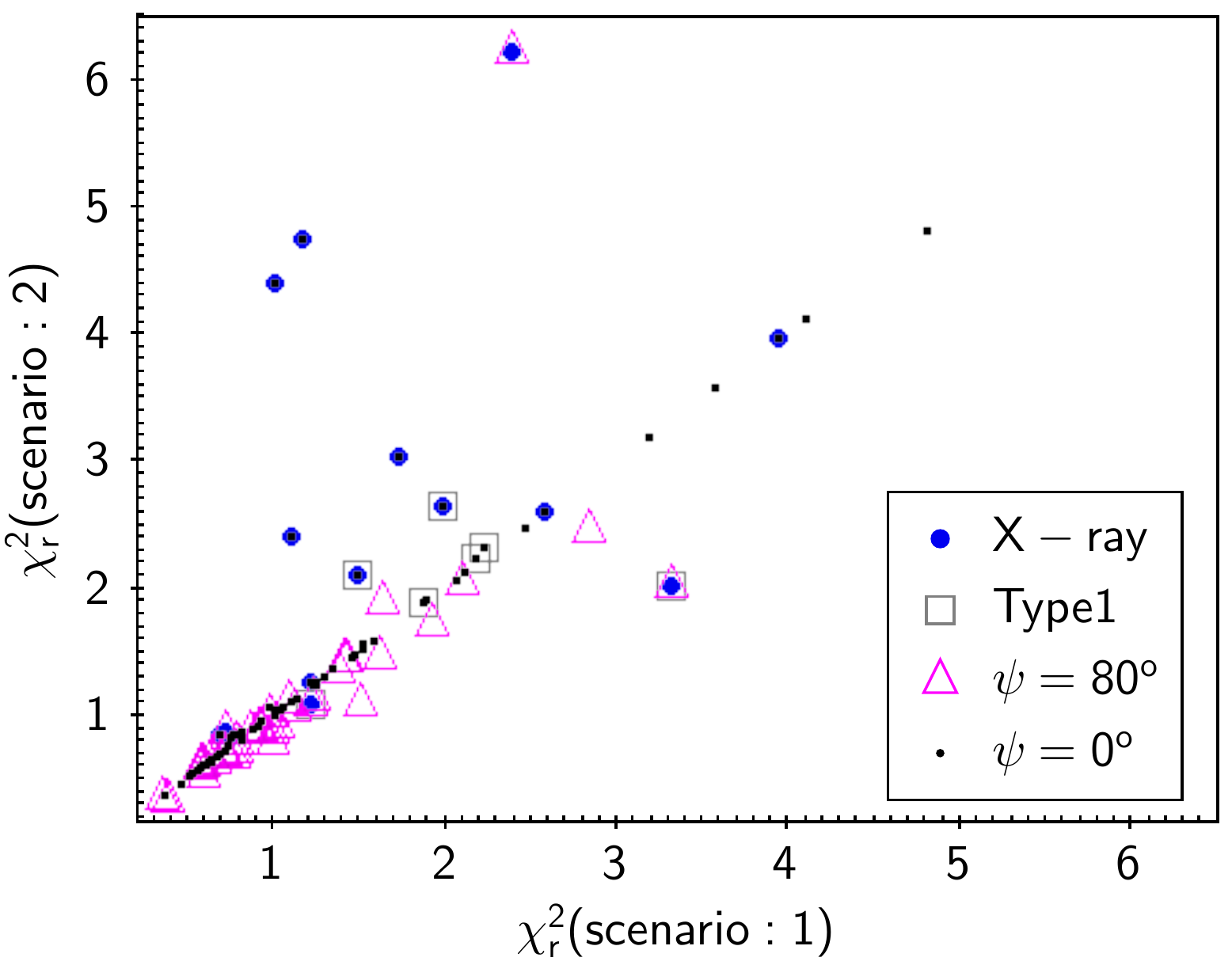}
   \includegraphics[width=9cm]{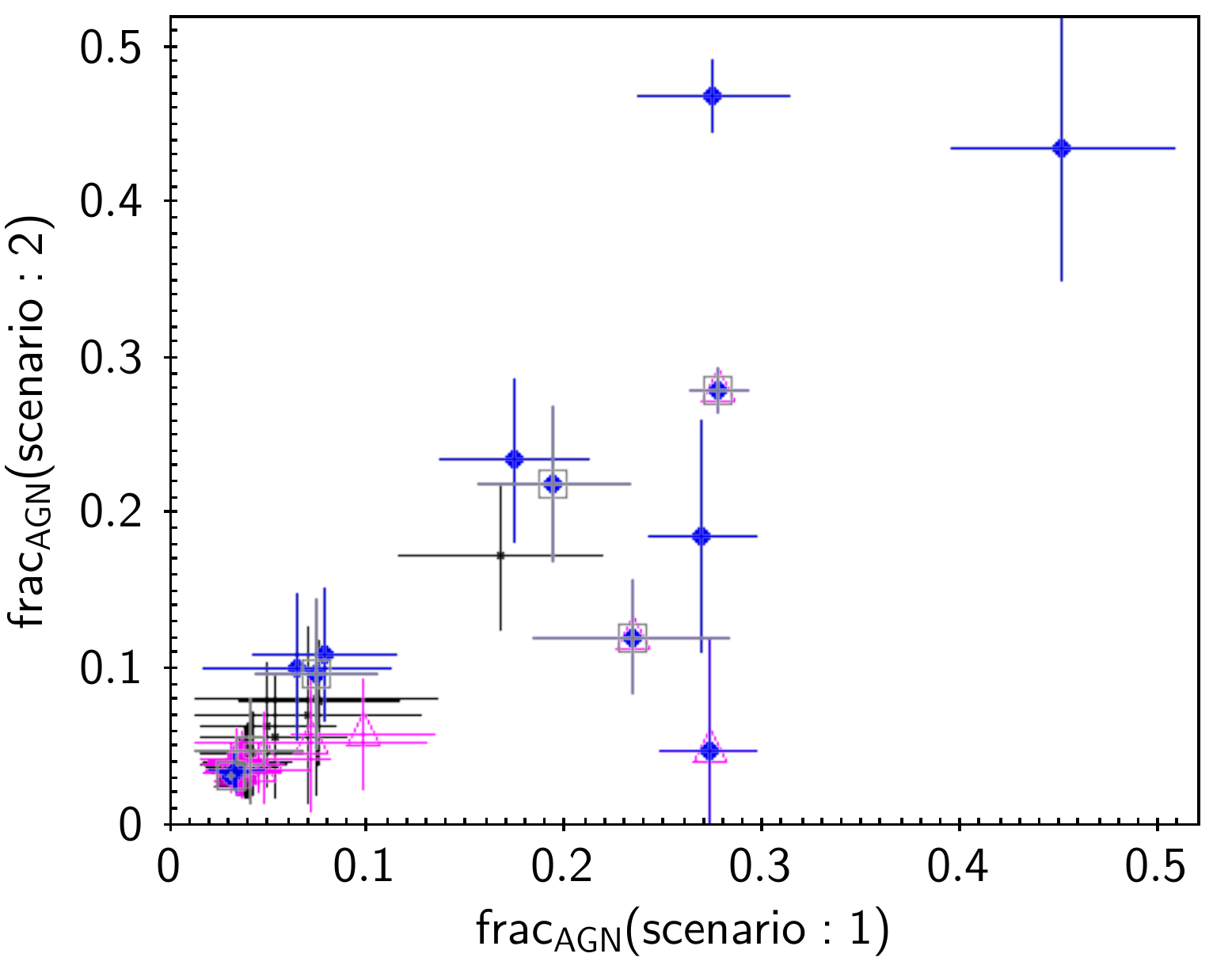}
   \caption{Comparison of $\chi_r^2$ and AGN fractions ($\rm frac_{AGN}$)  for the SPECZ sample. The results with scenario 1 for the AGN component ($\tau_{9.7}=1,6$ and $\rm \psi=0^o$)  are plotted on the x-axis and those for scenario 2  ($\tau_{9.7}=6$ and $\rm \psi=0,80^o$) on the y-axis.  Galaxies detected in X-ray, and galaxies spectroscopically classified as AGN type 1  are  plotted with blue points and empty squares, respectively.  The objects best fitted with either $\rm \psi=0^o$ or $\rm \psi=80^o$ for scenario 2 are differentiated  (black points and pink empty triangles).  The dispersion on the estimations of the AGN fraction (1 $\sigma$ dispersion of the PDFs) is plotted with  error bars. 
   }
              \label{agnfrac}%
    \end{figure}

\subsection{ Fitting sources with spectroscopic redshifts}

   \begin{figure*}
   \centering
\includegraphics[width=18cm]{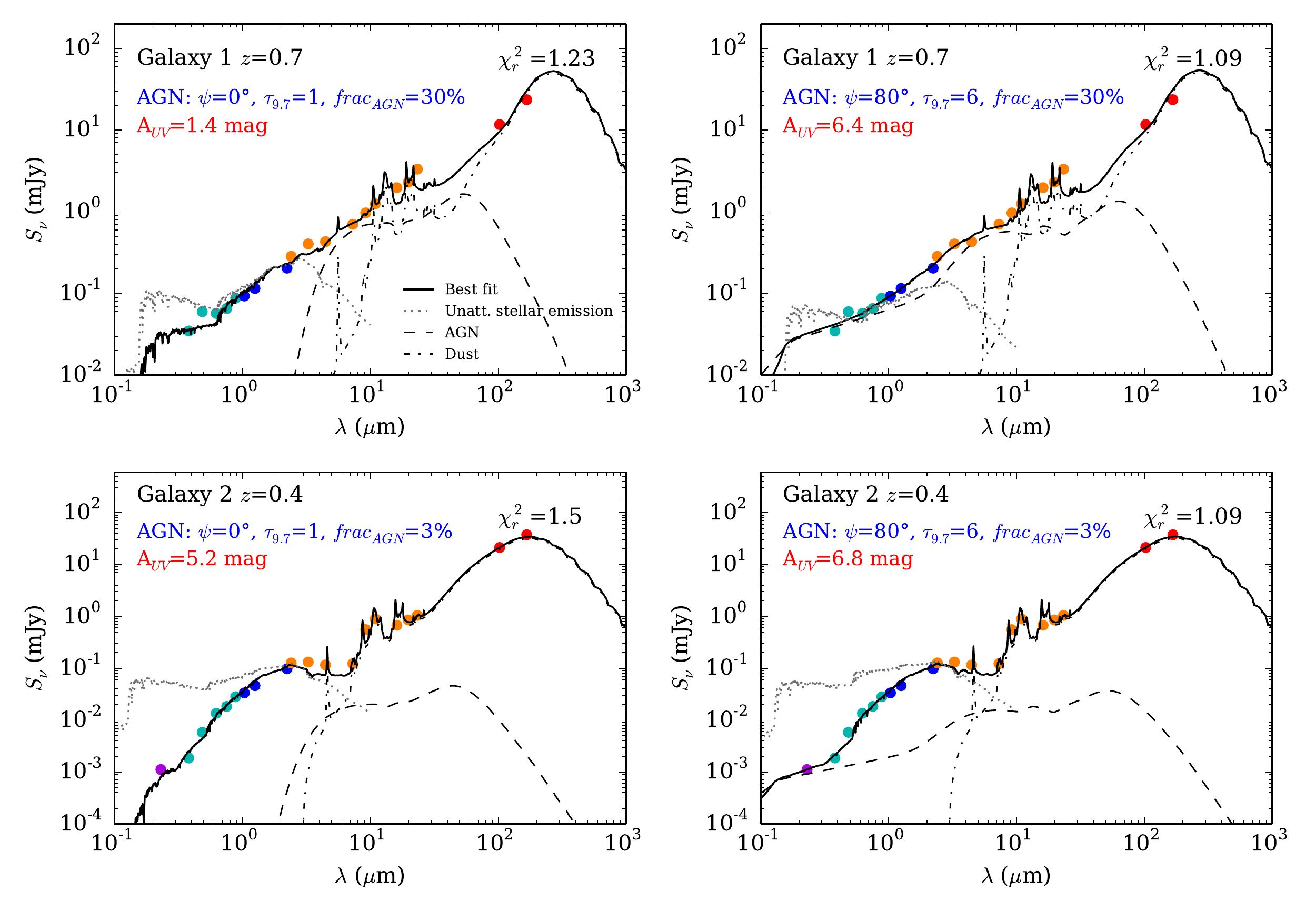}
   \caption{Examples of fits with the two scenarios for the AGN component. Symbols and lines are the same as in Fig. \ref{sed}. Galaxy 1 (upper panels) is detected in X-ray and identified as an AGN type 1,  galaxy 2 (lower panels) is a 'typical' galaxy of our sample, without any X-ray detection or spectral classification. The left panels correspond to  best models obtained with  scenario 1 ($\tau_{9.7}=1,6$, $\rm \psi=0^o$ ) and  right panels correspond to best models obtained with  scenario 2 ($\tau_{9.7}=6$ and $\rm \psi=0,80^o$). All  fits are good with consistent AGN fractions but the amount of dust attenuation varies according to the adopted scenario. }
              \label{agnfit}%
    \end{figure*}

The quality of the fitting process is checked on the SPECZ sample for which spectroscopic redshifts are available.  The fits are very good, 99 out of the 106 SEDs  fitted with a reduced $\chi_r^2$ lower than 3 and $<\chi_r^2>=1.3$. Some examples are given in Fig.\ref{sed}.

\subsubsection{AGN templates}
The SPECZ sample is well suited to check the influence of the choice of the AGN templates. We explore the two combinations of parameters presented above (scenario 1 and 2).
In Fig. \ref{agnfrac} we compare the reduced $\chi_r^2$ and AGN fractions found with these two configurations, and X-ray sources are checked separately. In the majority of cases a good agreement is   found for the $\chi_r^2$  and AGN fraction values.  
Only two values of $\tau_{9.7}$ or $\psi$  are introduced in both scenarios and the PDF of these parameters cannot be built: for the current discussion, we consider the values  of  $\tau_{9.7}$ and $\psi$ corresponding to the best model. When two input values of  $\tau_{9.7}$ are considered (scenario1, $\tau_{9.7}= 1,6 $ ), the best model corresponds to $\tau_{9.7}=1$ for 85 galaxies (80$\%$ of the cases).  When two values of $\psi$  are allowed (scenario 2, $\rm \psi=$0, $80^o$), $\psi$ is  found equal to $\rm 0^o$ (AGN type 2) for the best model  of 70 sources (66$\%$ of the cases).  
 Eight  galaxies exhibit a difference between their $\chi_r^2$ larger than 0.5. All of these galaxies  are detected in X-ray and three  are classified as AGN type 1 from their spectra.  Their AGN fractions are also found to be more dispersed than for the other galaxies. 
 In all but one case $\chi_r^2$ is significantly lower with  scenario 1. 
 Hereafter  we adopt  scenario 1, corresponding to AGN type 2 models with either a high silicate absorption or an almost flat mid-IR continuum, as our fiducial scenario for the AGN component.
 \\
It is worth noting that the choice of the AGN template substantially modifies its contribution to the UV-optical SED  as shown in Fig.\ref{agn}, and we expect some impact on the determination of the attenuation.
This impact   is illustrated in Fig.\ref{agnfit}  with  two examples:  a galaxy detected in X-ray and a  galaxy defined as 'typical' (without any X-ray emission or spectral classification). Their SEDs are fitted with  the two  scenarios  inducing  very different dust attenuations: 1.4 mag versus 6.4 mag for the X-ray source and 5.2 mag versus 6.8 mag for the 'typical' galaxy. AGN fractions are similar and    a low $\chi_r^2$ value is obtained for both scenarios.  The two corresponding best models are shown:  when only AGN type 2 templates  are considered (left panels, scenario 1)  the UV-optical SED is dominated by the stellar emission, whereas the quasar emission is found to dominate when the best model corresponds to an AGN type 1 (right panels, scenario 2).  
 Finally, we would like to emphasize the difficulty of fitting SEDs with multiple components. As shown in Fig.\ref{agnfit},  both fits  are satisfactory with $\chi_r^2 \le 1.5$ whereas the nature of the AGN component found for each best model (type 1 or 2) differs.

\subsubsection{Attenuation curves}
 \begin{figure}
   \centering
  \includegraphics[width=7cm]{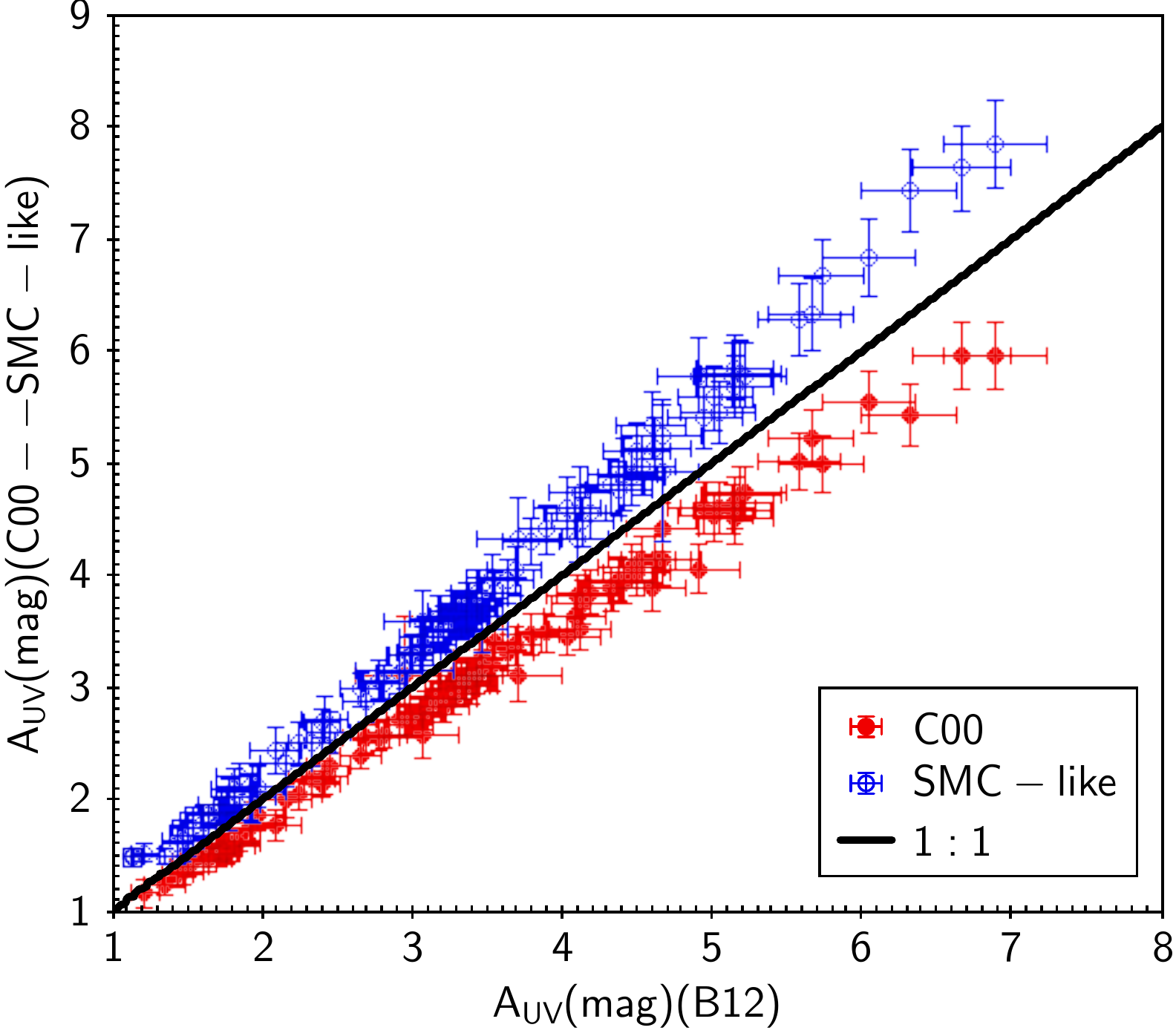}
   \includegraphics[width=7.2cm]{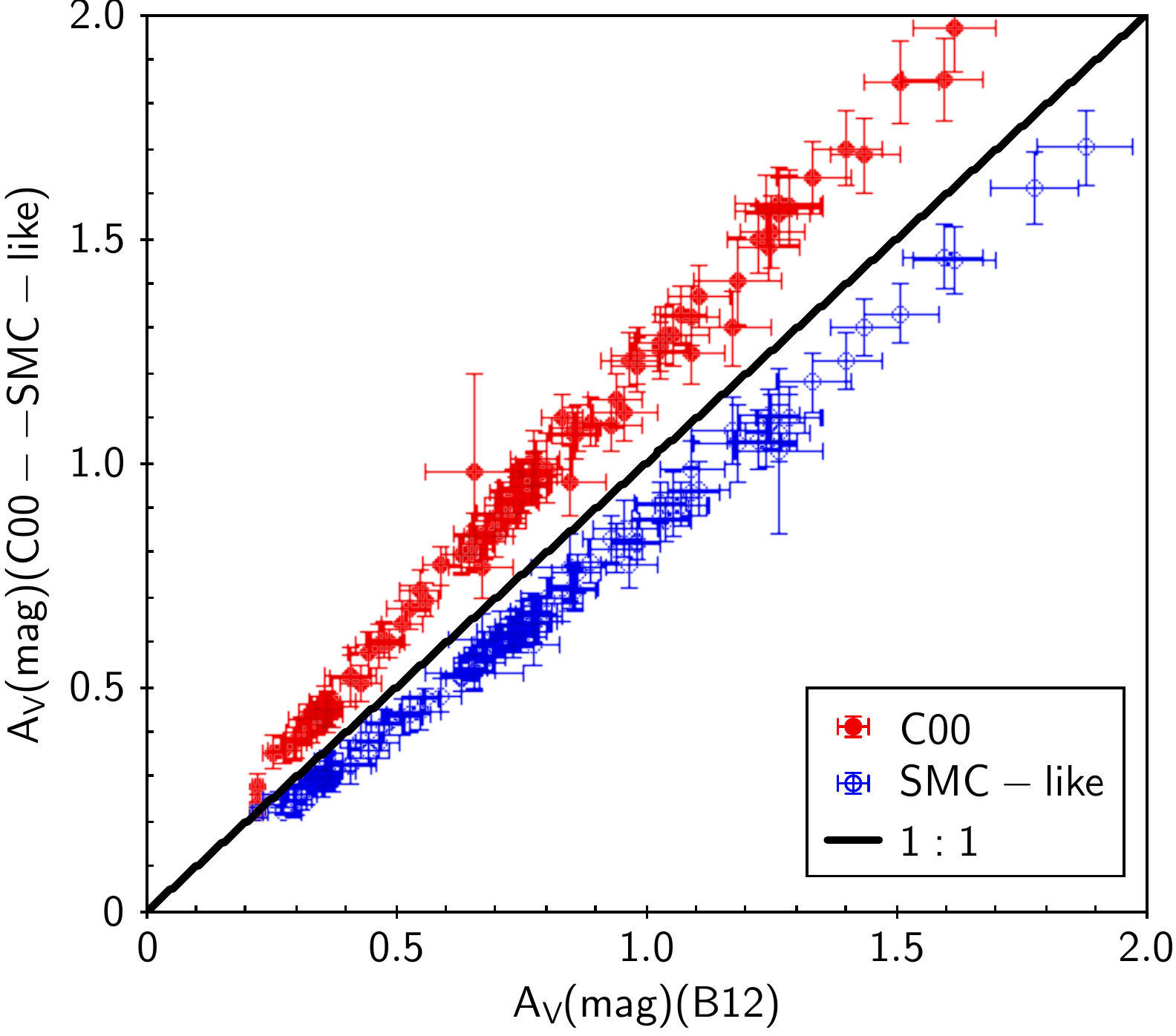}
   \caption{Comparison of attenuation factors in the UV (upper panel) and in the V band (lower panel) for the different attenuation curves. The values found for the B12 attenuation curve are plotted on the x-axis, and those  found for the C00 (filled red circles) and SMC-like (blue empty circles) attenuation curves are plotted on the y-axis. The dispersion of  the measures is reported as an error bar.}
              \label{attcurve}%
    \end{figure}

\begin{figure}
   \centering
  \includegraphics[width=7.5cm]{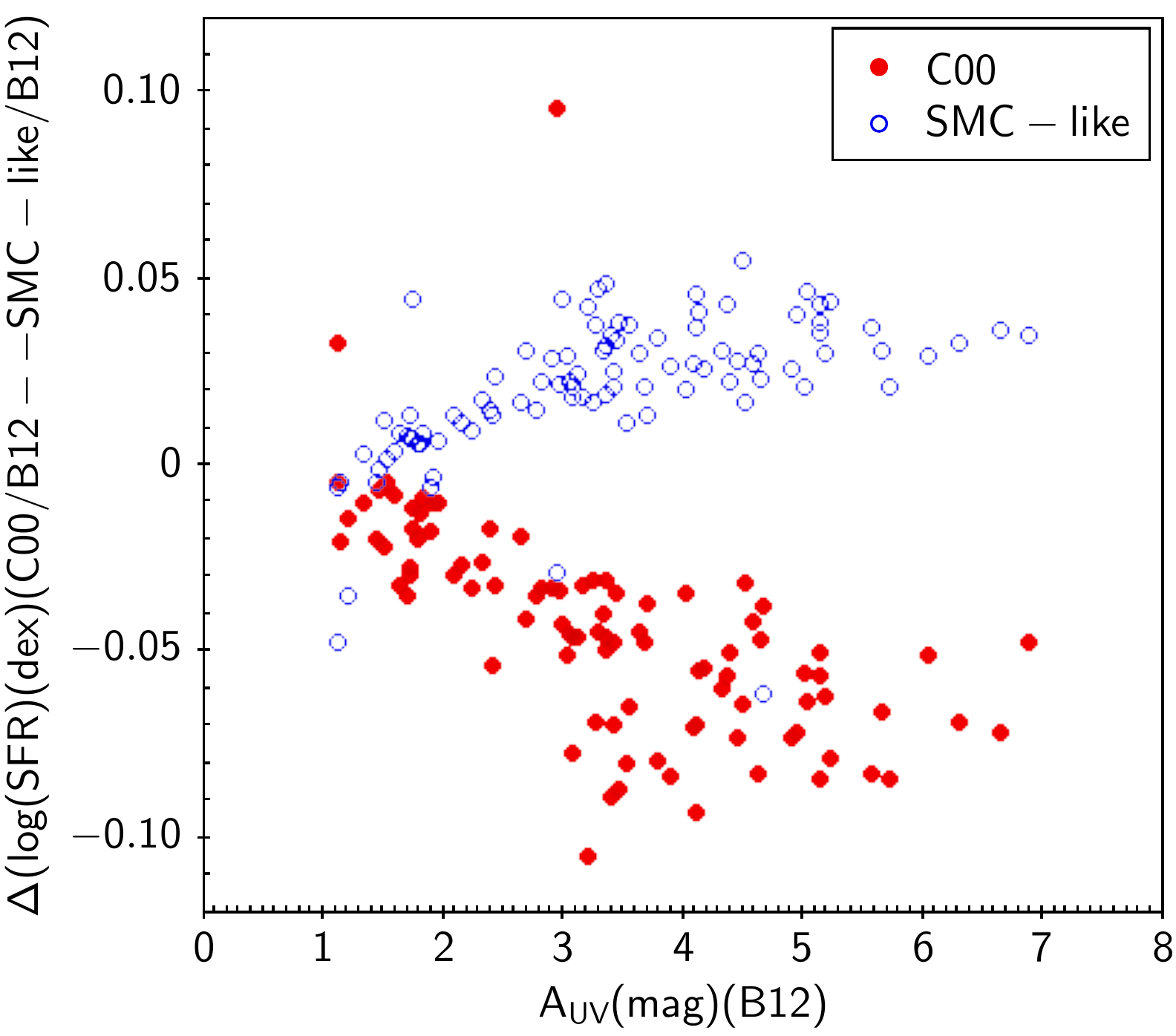}
   \includegraphics[width=7.5cm]{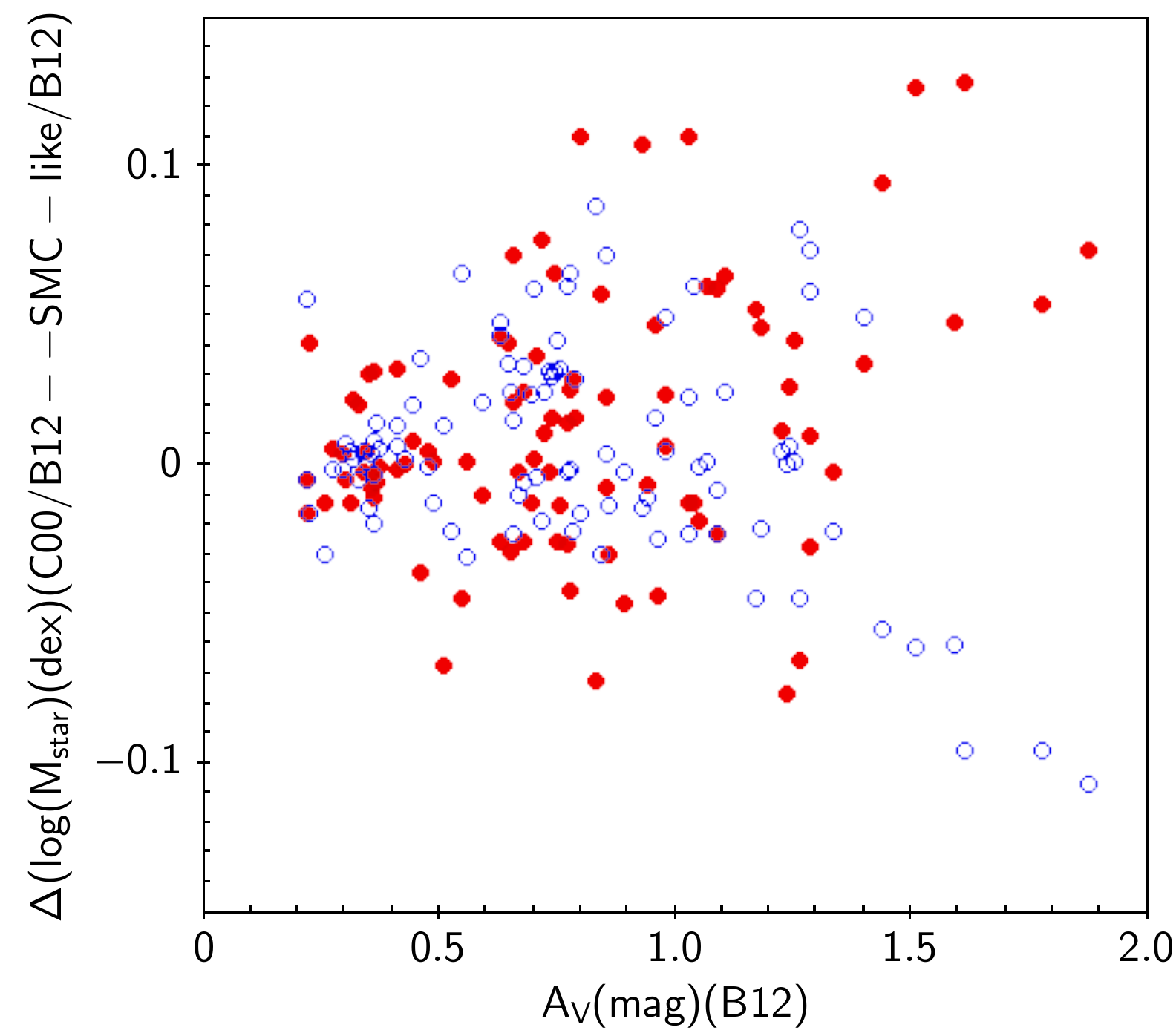}
   \caption{Comparison of SFR and stellar mass ($M_{\rm star}$)  estimations for different attenuations curves. The attenuation (in the UV  (resp. V band) for the SFR (resp. $M_{\rm star}$) estimations) is plotted on the x-axis. The difference (expressed with a logarithmic scale) between SFR and $M_{\rm star}$ estimations with the C00 and the SMC-like attenuation curves and the B12 attenuation curve is plotted on the y-axis. The symbols are the same as in Fig.\ref{attcurve}.}
              \label{sfr-mstar}%
    \end{figure}

Here we analyze   the impact of changing the  attenuation curve on the estimation of attenuation, SFR, and stellar masses. 
The code was run with the three different attenuation curves described in Sect. 3.2.1 (B12, C00 and SMC-like).  We adopt  scenario 1 as our fiducial scenario for the AGN component ($\tau_{9.7}=1,6$ and $\psi=0$) and check that the results do not depend on the assumed AGN templates (scenario 1 or 2). The fits are very good for the three attenuations curves: in all cases more than 90$\%$ of the fits correspond to $\chi_r^2 < 3$. The SMC-like curve is chosen as the best choice for 25 sources against 38  and 43 for C00 and B12, respectively, implying a slightly  lower pertinence for the SMC-like curve as compared to the two others.
Systematic differences are found for $A_{\rm UV}$ and  $A_{\rm V}$   as shown in Fig.\ref{attcurve}. These differences occur because of the limited amount of data (if any) in the UV rest-frame range. Using the B12 curve as  the reference, a larger attenuation is found in the V band for C00 together with a lower attenuation in the UV, and inverse trends are found for the SMC-like curve. Opposite variations found for the UV and V bands are explained by the shape of the attenuation curve. The C00 slope is flatter than the B12 and SMC-like slopes:  for the same amount of IR emission, the attenuation is lower in UV and higher in the V band. The SMC-like curve is the steepest, giving the highest attenuation in the V band and the lowest  in the UV.\\

Linear regressions between the different estimations  yield
\begin{align}
&A_{\rm UV-C00} = 0.89 \times A_{\rm UV-B12}+0.05,\\
&A_{\rm UV-SMC} = 1.12 \times A_{\rm UV-B12}-0.04,\\
&A_{\rm V-C00} = 1.18 \times A_{\rm V-B12}+0.04,\\
&A_{\rm V-SMC} = 0.89 \times A_{\rm V-B12}-0.03,
\end{align}
where the values obtained with the B12 curve ($A_{\rm V-B12}$ and $A_{\rm UV-B12}$) are taken as independent values for the regression.\\

The SFR is mainly constrained by the IR emission,  so we do not expect strong differences and only small variations are found. The variation of SFR is plotted against $A_{\rm UV}$ in Fig.\ref{sfr-mstar}.  $\rm <\Delta(log(SFR)>=-0.04$  dex  and $+0.02 $ dex  between C00 and B12, and SMC-like and B12, respectively. The discrepancy increases with attenuation as expected: the difference in attenuation increases with the attenuation itself. With  the C00  curve the UV attenuation is found to be  lower,  so is the SFR, the inverse holds for the SMC-like curve.\\

The estimations of stellar masses are unaffected by changing the attenuation curve as long as  $A_{\rm V} <\sim 1.4$ mag. For higher attenuations, when compared to the reference chosen here (B12 law), the C00 law gives larger stellar masses (corresponding to a larger dust correction in the V band) and the SMC-like curve  lower masses (corresponding to a lower correction). The differences remain lower than 0.1 dex on average (Fig.\ref{sfr-mstar}).

\

 \section{Evolution of dust attenuation}
 \begin{figure}
   \centering  
  \includegraphics[width=\columnwidth]{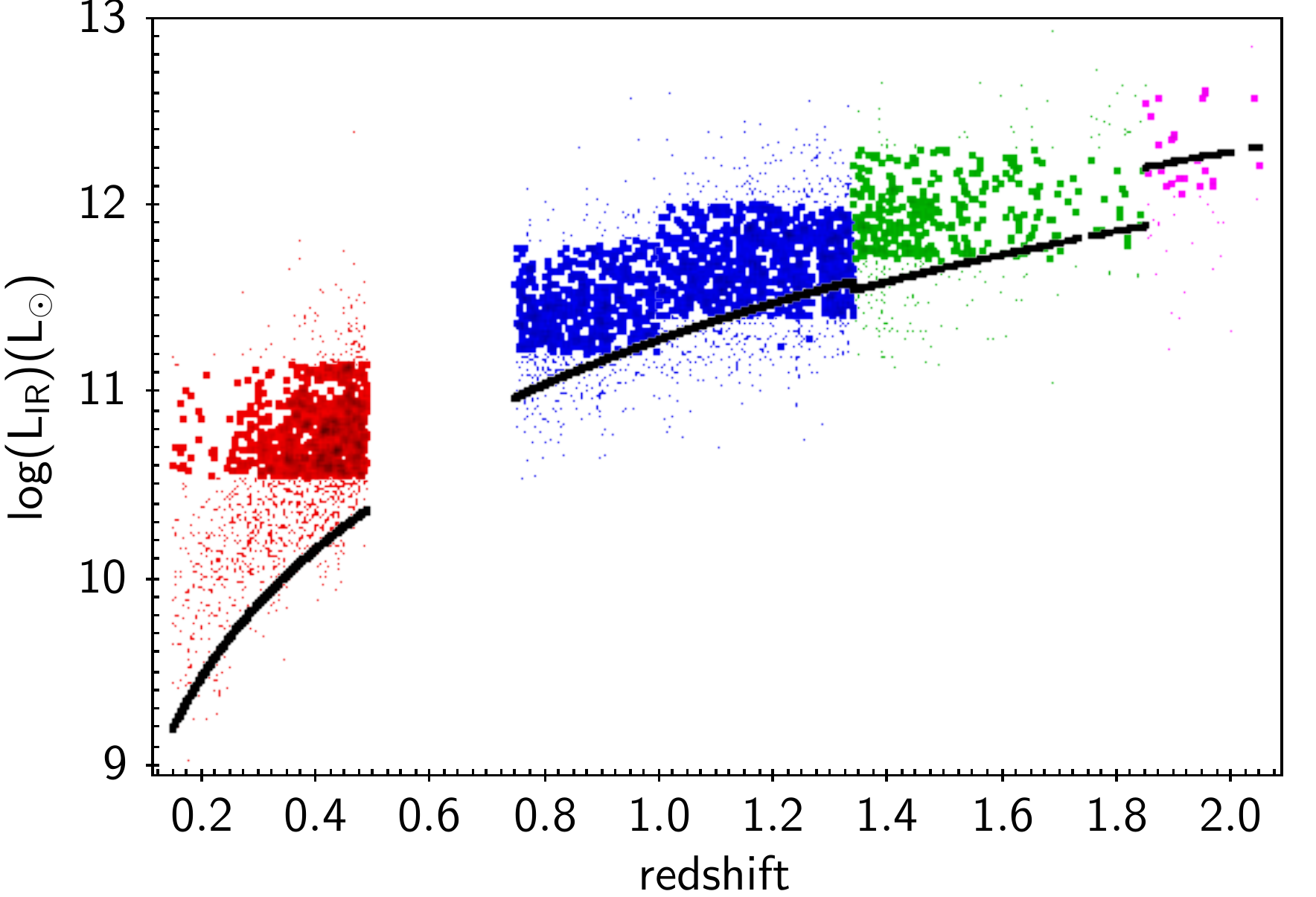}
   \caption{ Redshift and luminosity distributions of the sources. The total IR luminosity $L_{\rm IR}$, plotted on the y-axis, is an output of the SED fitting.  The small dots  represent the whole sample and the filled circles the subsample of galaxies selected around $L_{\rm IR}^*$ within each redsfhift bin (bin 1: red; bin 2: blue, split in two subbins corresponding  to $z<1$ and $z>1$; bin 3: green; bin 4: magenta). The adopted values of $\log(L_{\rm IR}^*/L_{\odot})$  10.84 (bin 1), 11.5 (bin 2, $z<1$), 11.7 (bin 2, $z>1$),  12 (bin 3), and 12.35 (bin 4). The bin size in luminosity  is 0.6 dex. Detection limits at 5$\sigma$  are reported as a black line.}
              \label{Ldust-z}%
    \end{figure}

\subsection{Definition of representative samples}

Our aim is to measure dust attenuation in galaxies selected with  similar criteria  at different redshifts.
The combination of the different IRC filters led us  to perform a  $8\mu$m rest-frame selection. In  Fig.\ref{Ldust-z} is reported the IR luminosity (obtained with CIGALE) against redshift for our sample. We  estimated the detection limit at 5$\sigma$ using the flux limits of \citet{murata13} for each IRC filter used to define the sample (cf. Table 1), and the average ratio between $L_{\rm IR}$ (found by SED fitting) and the monochromatic luminosity  in the IRC filter corresponding to the redshift selection.  Clearly, we cannot study  galaxies with a similar   $L_{\rm IR}$ over the full redshift range since the limit in luminosity increases sharply with z.  We adopt two strategies: 
  \begin{enumerate}
      \item {studying galaxies sampling the same domain of the luminosity function at  different redshifts,}
                 \item { selecting galaxies of similar luminosity on a reduced range of redshift.}
   \end{enumerate}
In the first case, we must    select the galaxies dominating the luminosity function and the luminosity density at a given redshift. To this aim, we take the total IR luminosity functions of \citet{magnelli13}, who fitted a double power-law function  originally  defined by \citet{sanders03}. In each redshift bin, we  select galaxies with an IR luminosity inside a bin of 0.6 dex centered on  the characteristic IR luminosity $L_{\rm IR}^*$ corresponding to  the transition luminosity of the double power-law function. The luminosity $L_{\rm IR}^*$  is calculated with  relations given by \citet{magnelli13}. The selection is  represented in Fig.\ref{Ldust-z}, and we split the bin 2 in two parts ($z<1$ and $z>1$). The selected sources lie above the 5$\sigma$ detection limit  for the  first three bins, except for a small overlap at the end of the second redshift bin. The detection limit in bin 4 does not allow us to sample the luminosity function well, and we drop this bin for this selection.\\
The average values of the stellar mass found for each redshift bin are reported in Table. 3. They are close to the specific values  found by \cite{ilbert10} for galaxies at similar redshift and with an  intermediate star formation activity. These values are also in the range of values found by \citet{karim11} for the galaxies dominating the star formation rate density up to $z=3$.\\

In the second approach, we want to measure the evolution of galaxies of similar $L_{\rm IR}$.  We define  a first luminosity bin, $11<\log(L_{\rm IR}/L_{\odot}) <11.4$, for galaxies with $z<1$ corresponding to the bin 1 and a part of bin 2 (Fig.\ref{Ldust-z}). At higher redshift, galaxies with $L_{IR}\ge 10^{11.7} L_{\odot}$ can be  observed in the redshift bins 2 and 3,  and we select galaxies in these bins with  $11.7<\log(L_{\rm IR}/L_{\odot}) <12.1$.

\subsection{SED fitting}

\begin{figure}
   \centering
   \includegraphics[width=\columnwidth]{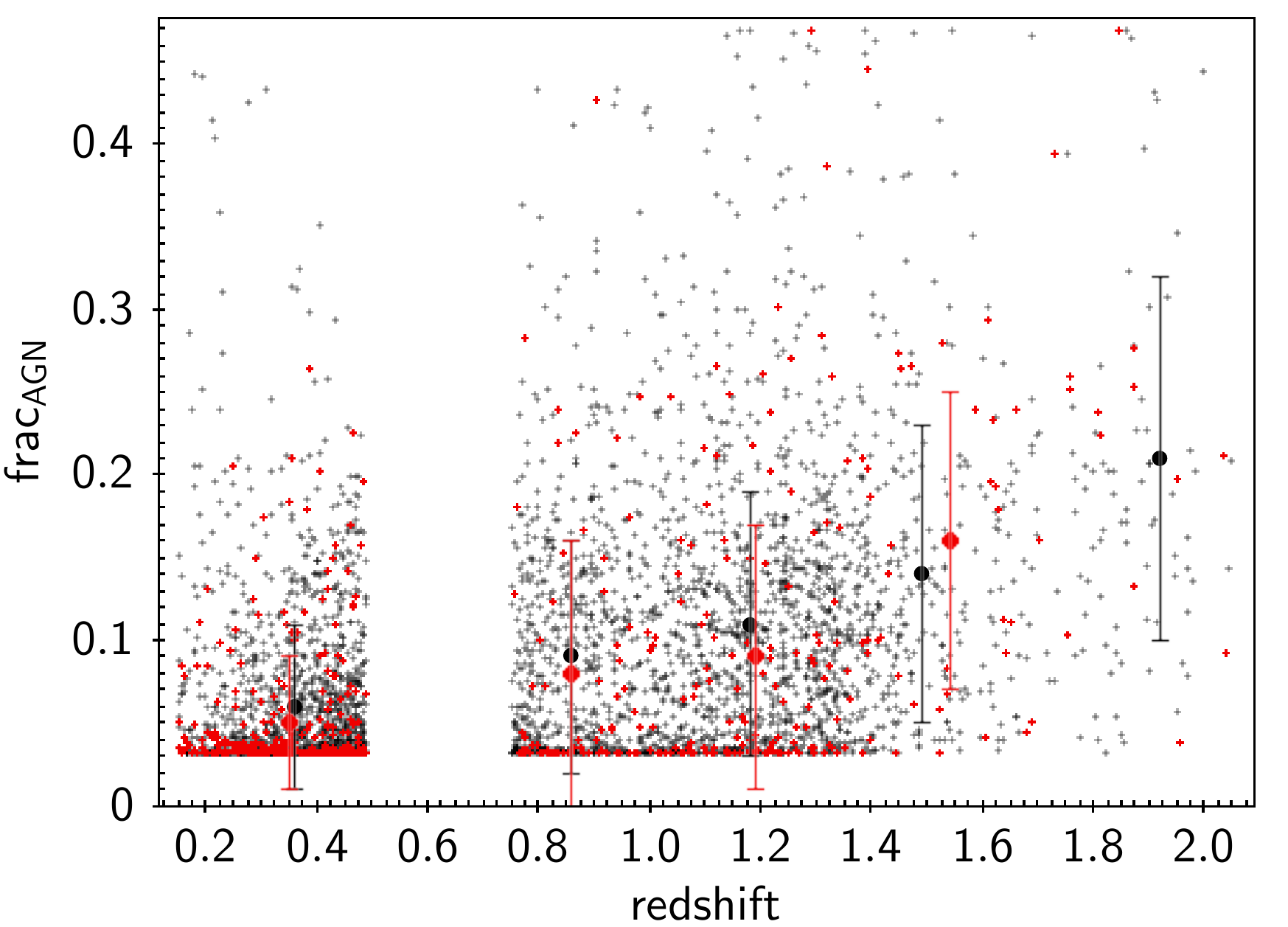}
   \caption{AGN fractions measured with CIGALE plotted versus the redshift of the sources: the whole sample is plotted with grey points, the measures for galaxies detected with PACS are represented with red points. The average values and dispersion for each redshift bin are over plotted (black filled circles for the whole sample, red filled circles for the PACS detections). The AGN fractions are always larger than 0 since they are measured as the mean of the PDF. The last redshift bin is not considered for the PACS detections  since it contains only 7 sources.}
              \label{AGN-z}%
    \end{figure}

CIGALE is run on the whole set of data (4077 sources) with a fiducial configuration (cf. Sect. 3). The reduced $\chi_r^2$ distributions are  very  good: for 93  $\%$ of the sources the minimum value of $\chi_r^2$ is lower than 5, with a similar result for each redshift bin and for galaxies detected with or without PACS.
We are studying dust attenuation in the UV with a sampling of this wavelength range that changes with redshift. We must check that our results are not biased by  varying wavelength coverage. We  have also considered  the subsample of galaxies with available UV rest-frame data that we defined in Sect. 2 and Table 1. The results for this subsample are very similar to those found for the whole sample.  \\
The AGN fraction is on average approximately $10\%$  for the whole sample  and is found to globally increase with z (Fig.\ref{AGN-z}).
Our spectral coverage  varies with redshift and, for most of the sources, we do not have  PACS data. With  only MIR  data, the decomposition  of the IR emission in a  stellar and a nonthermal component is difficult. We  checked that  the same trend is found when only galaxies detected with PACS are considered as shown in Fig.\ref{AGN-z} (as well as for our SPECZ sample, not shown here). The average value of $\rm frac_{AGN}$ is $6\pm5$, $9\pm 7$, $11\pm 8$,$14\pm8,$ and $21 \pm 10 \%$ from bin 1 to bin 4. Scenario 2 ($\tau_{9.7}=6$ and $\rm \psi=0,80^o$) yields slightly lower AGN fractions but also increases with redshift (6, 9, 11, and 15$\%$ from bin 1 to bin 4 with similar dispersions).  Our results are in global agreement with those of \citet{goto11}, who isolated AGN galaxies in their $AKARI$/SDSS sample  using  emission line ratios  and also found a very moderate contribution of AGNs to the global IR emission.

\subsection{Dust attenuation}
\subsubsection{Evolution of the amount of dust attenuation with redshift for $L_{\rm IR}^*$ galaxies}
\begin{figure*}
   \centering
\includegraphics[width=15truecm]{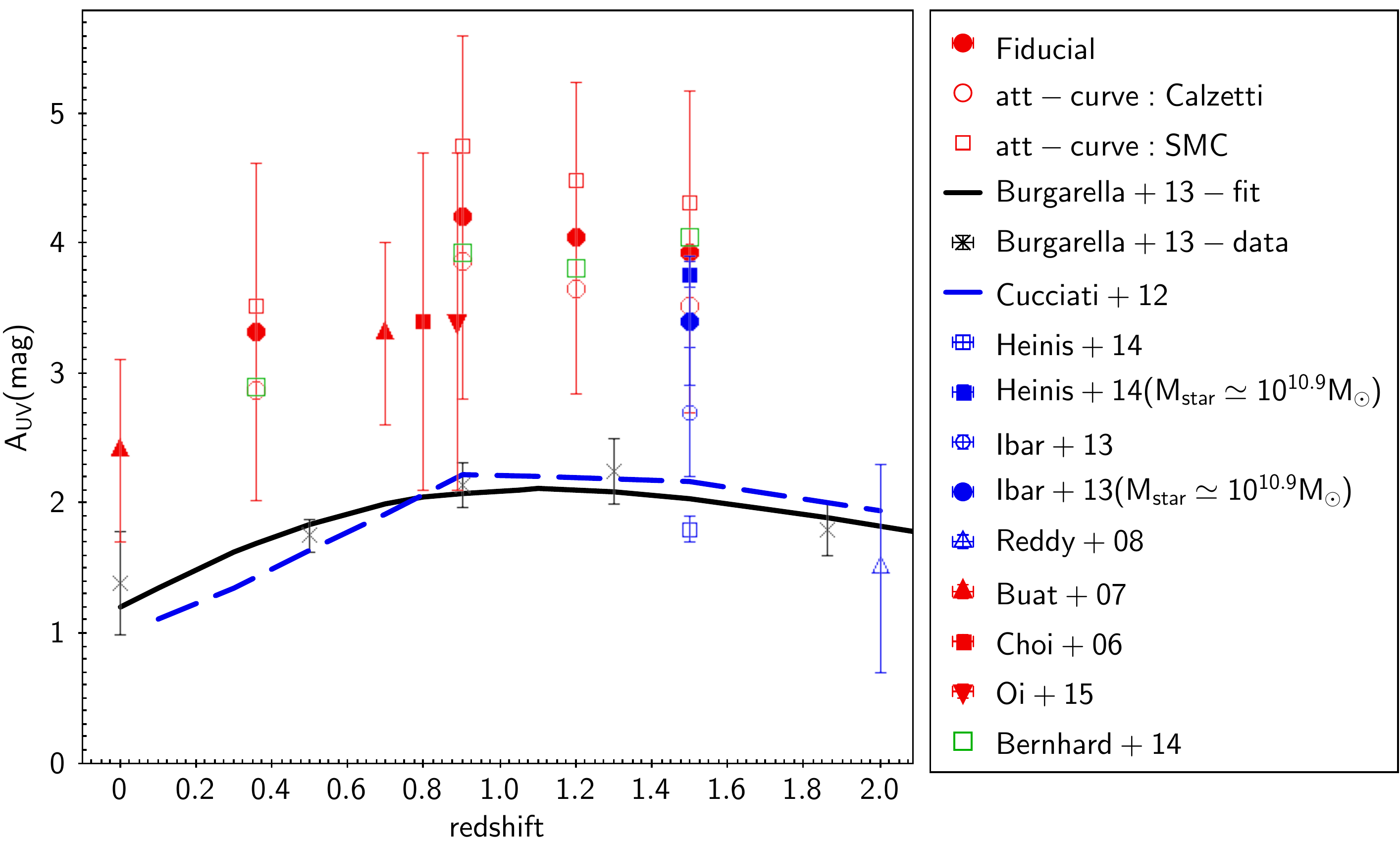}
 \caption{Average values of the attenuation factor, $A_{\rm UV}$,  versus redshift. The red symbols refer to  IR selected samples. The results of the present study are plotted with red-filled circles for the fiducial modeling, empty circles and squares are used for the C00 and SMC-like attenuation laws. Measurements from \citet{buat06,buat08} are plotted with red-filled triangles, those from \cite{choi06} and Oi et al.  with a filled red square and a filled red triangle. The blue symbols and lines refer to   UV and $\rm H_{\alpha}$ line selections discussed in Sect. 5. \cite{cucciati12} results are reported with a   blue-dashed line. The blue squares refer to measurements from \citet{heinis14} based on a stacking analysis: the empty square for the whole UV selection and the filled square  for galaxies with $M_{\rm star}=10^{10.9} M_{\odot}$. The blue circles refer to measurements from \citet{ibar13} for $\rm H_{\alpha}$ emitters: the empty circle is  for the whole selection and filled circle for galaxies with $M_{\rm star}=10^{10.9} M_{\odot}$. The blue empty triangle refers to measurements from  \citet{reddy08}. The black line and crosses correspond to  the global estimates of   \cite{burgarella13} based on luminosity functions. Their analytical relation is plotted as a solid line and  crosses correspond to  their  measurements at different redshifts.   The  dispersion of each measurement is reported with a vertical bar. The measure of the dispersion for the \cite{heinis14} values are based on a stacking analysis whereas the others are estimated based on individual detections. The green empty squares are the average values of the attenuation found with the model of \citet{bernhard14}, discussed in Sect.5.}
              \label{Afuv-z}%
    \end{figure*}

To study the average  attenuation as a function of  redshift, we  first consider  the  subsample of galaxies with luminosities around $L_{\rm IR}^*$.   The average values of $\rm A_{UV}$ for this selection are reported in Table 3.    \\
We show in Sect.3 that the measure of the  attenuation  depends on the choice of the attenuation curve. We  can calculate the average values of the attenuation in each redshift bin for the different attenuation curves that we consider  by applying  regression relations (Eq. 3 and 4).  The variation with redshift of the different estimations of the  attenuation is shown in Fig.\ref{Afuv-z}. \\
 We must check that the definition of $L_{\rm IR}^*$  as well as the  expected increase of the uncertainty on  the measure of $\rm A_{UV}$ with the  AGN fraction do not affect our results.   To this aim, we also  consider  the whole galaxy sample in each redshift bin and  the sources with an AGN fraction lower than 10$\%$. With  the whole galaxy sample the average values of $\rm A_{UV}$ are found 0.2 mag lower than for the subsample of galaxies with luminosities around $L_{\rm IR}^*$ except for the first bin for which the difference reaches 0.3 mag (as expected since we add a large number of fainter galaxies). When only galaxies with an AGN fraction lower than 10$\%$ are considered, the difference never exceeds 0.2 mag. These variations are lower than  the difference induced for example by changing of the attenuation law (Fig.\ref{Afuv-z}).

We can add  other measures of attenuation, already obtained for IR selected galaxies, to our analysis. At $z=0$, \citet{buat06} combined $GALEX$ and $IRAS$ data and derived volume averaged measures of the attenuation. For   a luminosity  $L_{\rm IR}^*\sim 10^{10.5} L_{\odot}$ \citep{sanders03}, the average UV attenuation  $A_{\rm UV} = 2.4\pm 0.7$ mag. \citet{buat08} measured the attenuation of luminous IR Galaxies (LIRGs, $L_{\rm IR} > 10^{11} {\rm L_{\odot}}$) at $z=0.7$ by combining 
{\it Spitzer} and $GALEX$ data and found a mean attenuation of 3.33 mag with a dispersion similar to that  found at $z=0$.  \cite{choi06} also measured dust attenuation  in mid-IR selected galaxies of the {\it Spitzer} First Look Survey at  $z=0.8$ by comparing SFR measured with the strength of emission lines and $L_{\rm IR}$.  We apply their relation between $A_{\rm V}$ and $L_{\rm IR}$ to the average value of $L_{\rm IR}$ for the redshift bin 2 (Table 3) and get $A_{\rm V}=2.33$ mag. The visual extinction  can then be  translated to an attenuation in the UV continuum   as explained in \cite{buat08} giving $A_{\rm UV} = 3.4$ mag. The dispersion is directly measured on Fig.12  of \cite{choi06}. Oi et al. (in preparation) observe the H$\alpha$ emission line of a sample of mid-IR bright sources at $z=0.9$  with the Subaru/FMOS spectrograph. We consider their 25 confirmed detections: the  average attenuation $A_{\rm H_{\alpha}} = 1.5\pm 0.60$ mag, which translates to $A_{\rm UV}= 3.4\pm 1.3$ mag (using the relations detailed in Sect.5.2).\\
All these  measures are overplotted on  Fig.\ref{Afuv-z}. Given the different methods applied to measure $A_{\rm UV}$ they are found to be consistent  and our obtained values      extend the analysis  up to $z\sim 1.5$.  The attenuation in UV for $L_{\rm IR}^*$ galaxies  is found to increase  with redshift from $\sim 2.5$ mag at $z=0$ to $\sim 4$ mag at $z\simeq 1$ and then to remain almost constant up to $z=1.5$.\\
\begin{table}
\caption {Evolution of dust attenuation at UV wavelength  with redshift for the fiducial model (B12 attenuation curve and AGN 2) and galaxies with a luminosity close to $L_{\rm IR}^*$. Mean values of redshift, $L_{\rm IR}$, $M_{\rm star}$ , and $A_{\rm UV}$  as well as the number N of sources used are reported for each redshift bin (bin 2 is split in two subbins, see text).  The dispersion of $A_{\rm UV}$ is on the order of 1.3 mag in each bin   (as plotted in Fig.\ref{Afuv-z}). }
\begin{tabular}{cccccc} 
\hline
 Bin & N & $< z>$ & $<\log(L_{\rm IR})>$ & $<\log(M_{\rm star})>$&$<A_{\rm UV}>$\\
 & &&$\rm L_{\odot}$& $\rm M_{\odot}$&mag\\

               \hline
  1 & 795 & 0.4&10.78 &10.41&3.32\\
  2-1 & 479  & 0.9&11.46 &10.61&4.20\\
  2.2 &726 & 1.2&11.68 &10.73&4.04\\
  3 & 282 & 1.5&11.96 &10.96&3.93\\
 
 \hline
 \end{tabular}
\end{table}

\subsubsection{Evolution of the amount of dust attenuation with redshift for galaxies of fixed $L_{\rm IR}$}

\begin{figure}
   \centering
  \includegraphics[width=\columnwidth]{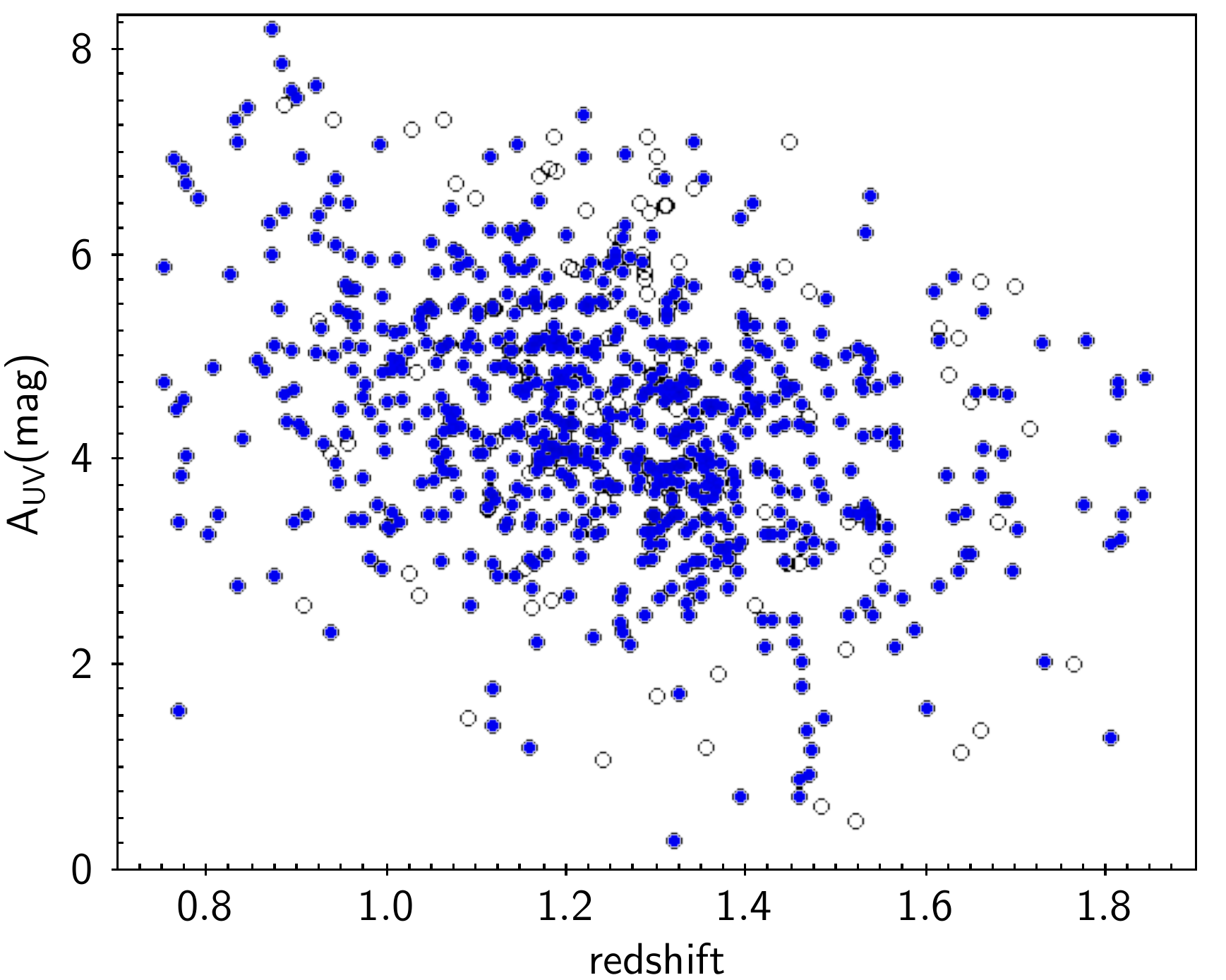}
 
    \caption{Attenuation factor, $A_{\rm UV}$, versus redshift for individual galaxies with $11.7< \log(L_{\rm IR}/L_{\odot})<12.1$. Measurements for galaxies detected in their UV rest-frame are plotted with blue filled circles, the empty black circles correspond to galaxies without UV rest-frame data.}
    \label{ir117}
    \end{figure}

We  also have access to  the amount of dust attenuation in  galaxies of similar $L_{\rm IR}$ on limited ranges of redshifts. For galaxies with $11<\log(L_{\rm IR}/L_{\odot}) <11.4$  and $z<1$,  we find $<A_{\rm UV}>=4$ mag at $<z>=0.4$ (bin 1) and $<A_{\rm UV}>=3.6$ mag at $<z>=0.9$ (bin 2 with $z<1$). The dispersion in both cases reaches 1.3 mag. For comparison, we can add  the results of \citet{buat08} for LIRGs with $<A_{\rm UV}>=3.8$ mag at $z=0$ and 3.3 mag at $z=0.7$ with a dispersion about 0.7 mag. The results are consistent given the dispersion of the measurements and both show a slight decrease of the attenuation when z increases.\\
In the redshift bins 2 and 3, we select galaxies with  $11.7<\log(L_{\rm IR}/L_{\odot}) <12.1$.  The variation of $A_{\rm UV}$ with z for these objects is plotted n Fig.\ref{ir117}.  Despite a very large dispersion, a decrease of  attenuation with z is found, the trend is  also clearly found when only galaxies detected in their UV rest frame are considered.

    \subsubsection{Variation of the amount of dust attenuation with $L_{\rm IR}$ and $M_{\rm star}$}
\begin{figure}
   \centering
  \includegraphics[width=\columnwidth]{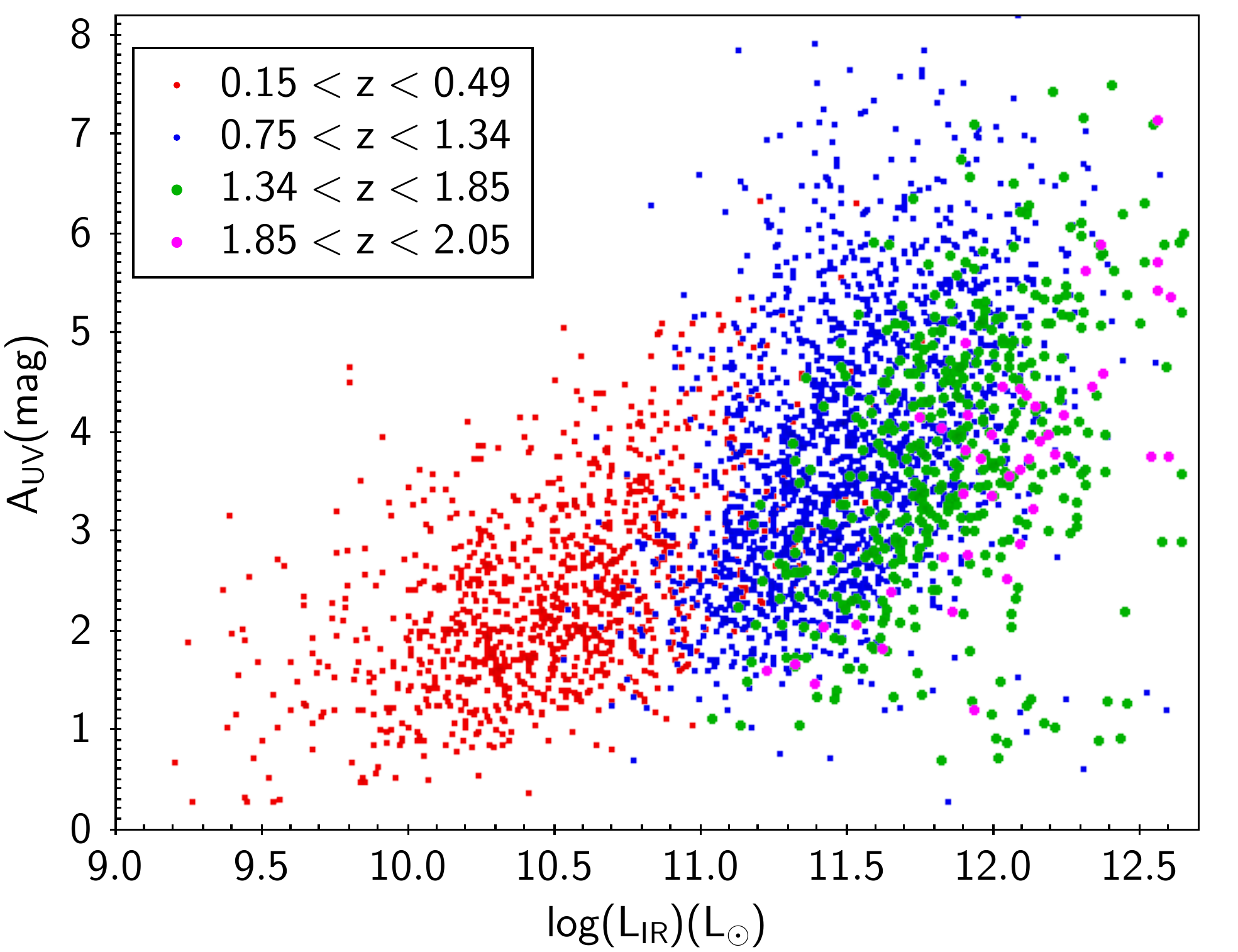}
    \includegraphics[width=\columnwidth]{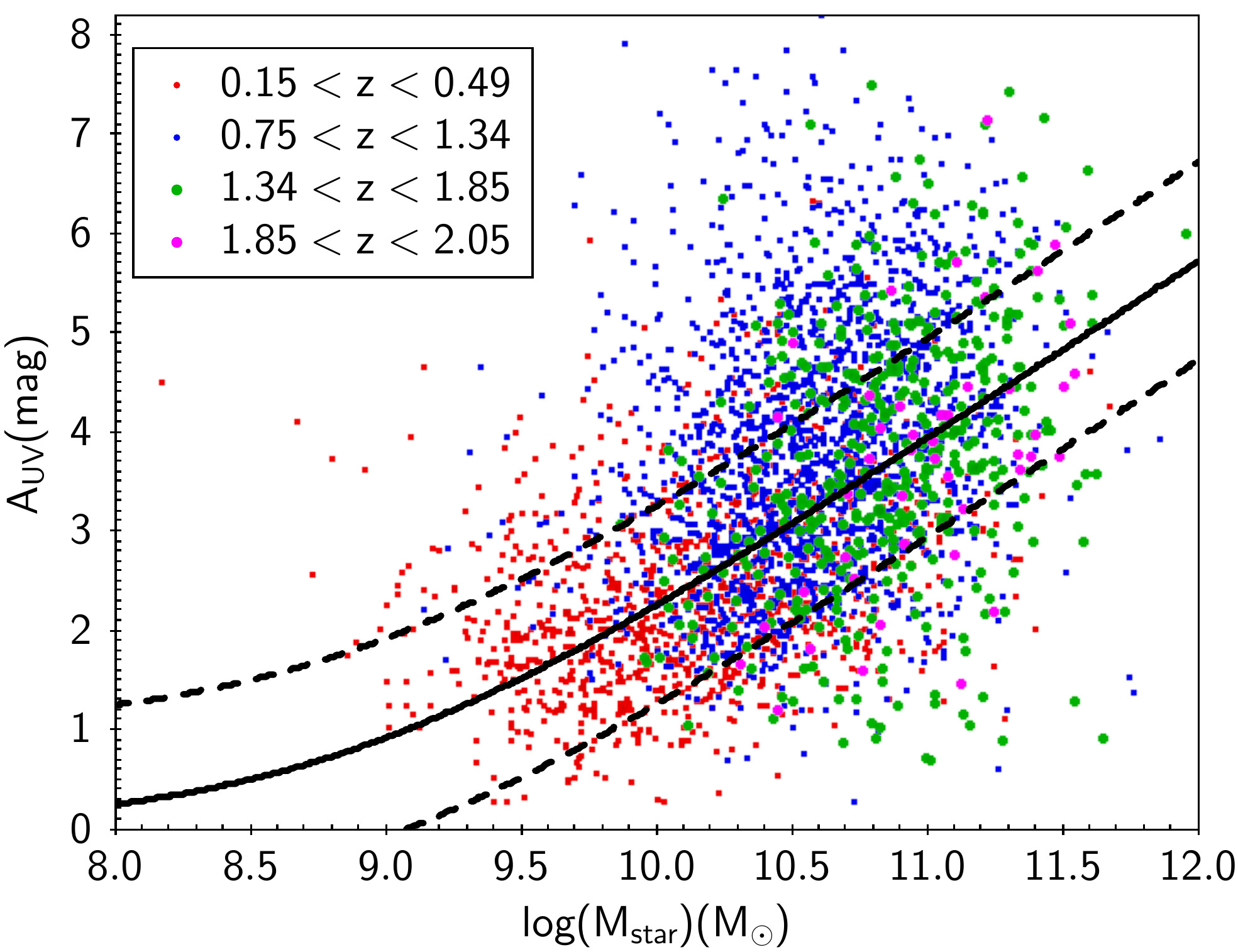}
    \caption{Attenuation factor, $A_{\rm UV}$,  versus $L_{\rm IR}$ (upper panel) and $M_{\rm star}$ (lower panel). The color coding is the same as in Fig.\ref{Ldust-z}, larger symbols are used for  bins 3 and 4 to improve their visibility. In the lower panel,  the relation of \citet{heinis14}  is plotted with a black solid line, the dotted line is the 1$\sigma$ dispersion of the relation (see text for details).}
    \label{auv-lir}
    \end{figure}

   Many previous studies have concluded that dust attenuation  increases with SFR or bolometric emission of young stars \citep[e.g.,][]{hopkins01, martin05,choi06,xu06,buat06,zheng07} and stellar mass \citep[e.g.,][and reference therein]{brinchmann04,martin07,iglesias07,pannella09,garn10,buat12}. Recent studies find that the relation between attenuation and stellar mass ($M_{\rm star}$) should not evolve with redshift \citep{ibar13,heinis14,pannella14}. 
In Fig.\ref{auv-lir},  $A_{\rm UV}$ is plotted versus $L_{\rm IR}$ and  $M_{\rm star}$ for our complete galaxy sample since there is no need here to restrict the analysis to any subsample. A general increase of $A_{\rm UV}$ with $L_{\rm IR}$ is found that explains the increase of the average attenuation with redshift  for galaxies close to $L_{\rm IR}^*$. The very large dispersion leaves room for    the slight decrease found with z for galaxies with a fixed $L_{\rm IR}$. \\
When $A_{\rm UV}$  is plotted against $M_{\rm star}$ the  redshift bins appear to overlap, except in the case of the first bin,  which samples much lower masses than the others. No clear trend is found  with z in agreement  with a nonevolution  with redshift at least for $z>0.5$. There is a global increase of $A_{\rm UV}$ with $M_{\rm star}$  with a very large dispersion.  In  Fig.\ref{auv-lir} we overplot the relation between $A_{\rm UV}$ and  $M_{\rm star}$ found by \cite{heinis14} with a dispersion of 1 mag  measured by \citet{pannella14} using $Herschel$ detections.  Half of the sample, lying between the limits of the relation at $\pm 1 \sigma$ and 35 $\%,$ are found above the upper limit, which roughly correspond to an excess of 20$\%$ of galaxies above the $A_{\rm UV}$-$M_{\rm star}$ relation of  \citet{heinis14} and \citet{pannella14}.

 \section{Discussion}

We now compare the redshift evolution of the attenuation measured with different methods and in samples selected with different criteria. We consider the global measure of \citet{burgarella13} based on luminosity densities  as well as   measurements performed in  large samples of star-forming galaxies selected to be either UV or H$\alpha$ emitters. We begin by a discussion of  these  measurements  and then we compare them to our results for our mid-IR selection.\\
\subsection{Average dust attenuation in the universe and in  UV selected samples}
 \citet{burgarella13} measured the global attenuation in the universe by comparing the IR and UV luminosity densities. Their result is plotted in Fig.\ref{Afuv-z}  (black line and crosses). This measure is not related to individual objects and can be considered as a measure of the attenuation averaged over all the galaxy populations.  This measure is much lower than the one we measure in our sample selected in IR. The difference is on the order of 2 mag (1.6 mag for  bin 1).\\

 \cite{cucciati12}   selected galaxies in their UV rest frame in the VVDS-0226-04 field up to $z=4.5$. They derived dust attenuation for each galaxy of their UV selection through SED fitting (without IR data), the variation they found is in close agreement with the results of \citet{burgarella13} as shown in Fig.\ref{Afuv-z}. \cite{heinis13} measured the average attenuation of a UV selection at $z\sim 1.5$ in the COSMOS field by stacking $Herschel$/SPIRE images. The average value found for their whole selection is reported in Fig.\ref{Afuv-z} and  is consistent with that derived by \cite{cucciati12} at the same redshift. In Fig.\ref{Afuv-z} we also report the measure of \cite{reddy08} at $z\sim 2$ for a sample of galaxies selected on their UV rest-frame colors,  which also agree with the other measurements.  The amount of attenuation found for our IR selection is much higher than the values found either in a UV selection or globally in the universe.\\

  \begin{figure}
 \centering
\includegraphics[width=\columnwidth]{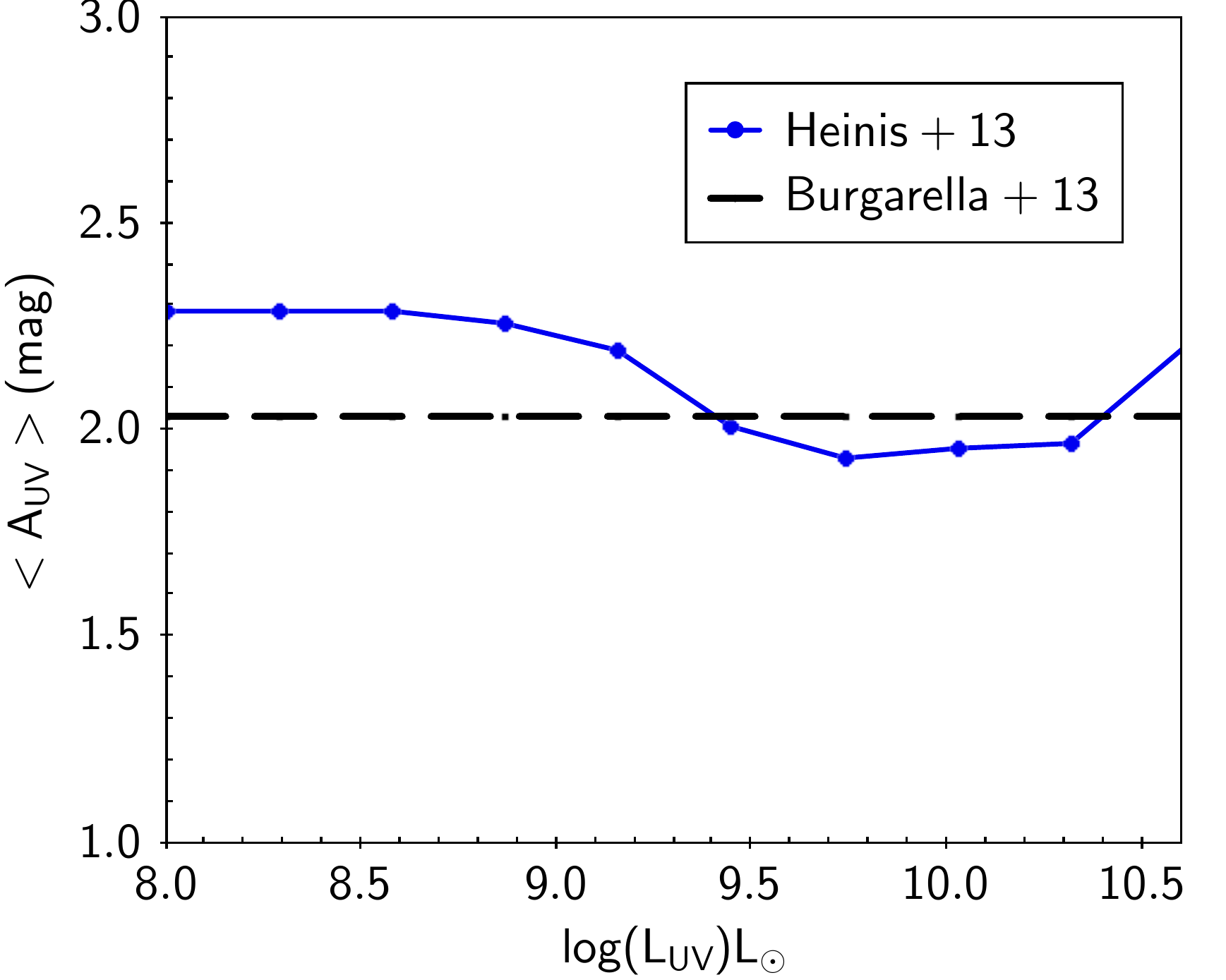}
  \caption{Prediction of the average amount of attenuation  in a UV selected sample at z=1.5 from the model of \citet{heinis13}. The threshold in  $L_{\rm UV}$ (observed)  is plotted on the x-axis and the  corresponding average  amount of dust attenuation $A_{\rm UV}$ is plotted on the y-axis with a blue line (see text for details). The dotted black horizontal line is the measure of \citet{burgarella13} at the same redshift.}
\label{IRX-uvsel}
\end{figure}

At first glance the agreement between \cite{burgarella13} and \citet{cucciati12} or \citet{reddy08} measurements may be surprising since they are obtained using very different approaches. Cucciati et al.   measured the attenuation in individual galaxies whereas the values reported by \citet{burgarella13} and based on luminosity functions are averaged over the galaxy populations and cannot be directly compared to individual measurements. \\
 These similar results can be explained by the lack of correlation between the average attenuation and  $L_{\rm UV}$ observed,  obtained   by  \citet{heinis13}.  We use their simulated catalog  built to reproduce the IR properties of their UV selected sample at $z=1.5$: we cut it at different values of $L_{\rm UV}$ and calculate the  luminosity density in UV and  in IR of the selected objects. The ratio of these luminosity densities is then translated in $A_{\rm UV}$ as in \citet{burgarella13} for comparison purposes. The resulting attenuation is plotted against the threshold in  $L_{\rm UV}$  in Fig.\ref{IRX-uvsel} and compared to the value found by  \cite{burgarella13}  at the same redshift. It is clearly seen that the cut in $L_{\rm UV}$ has no significant  impact on the average measure of the attenuation.
Therefore a galaxy sample selected in UV, as in \citet{cucciati12} or \cite{reddy08},  can be   representative of the average  attenuation in the universe, without systematic bias due to a limiting UV luminosity.  
\subsection{Dust attenuation of H$\alpha$ emitters}

\citet{ibar13} analyzed   the IR properties of galaxies detected in their H$\alpha$ line (HIZELS project) at $z=1.46$. From a stacking analysis of {\it Spitzer}, {\it Herschel,} and AzTEC images  they derived the average total IR luminosity of their sample. Using a local calibration between H$\alpha$  line and IR continuum emissions from \citet{kennicutt09} they obtained  a median attenuation $A_{\rm H\alpha}=1.2 \pm 0.2$ mag. This attenuation can be translated in UV using the recipe of \cite{calzetti97}. The attenuation $A_{\rm H\alpha}$ is related to the color excess of the ionized gas  by applying a standard  extinction curve,
i.e., \begin{equation}
E_g(B-V) = A_{\rm H\alpha}/2.45$$ 
,\end{equation}
adopting a Galactic extinction curve  \citep{pei92} and a screen geometry.

The color excess of the stellar continuum $E (B-V)$  is  related to the color excess of the gas   \citep{calzetti97}, i.e.,
\begin{equation}
E (B-V) = 0.44 \times E_g (B-V)
.\end{equation}
The attenuation in UV  is then calculated with  the adopted attenuation law. To be consistent with our analysis,  we use  the B12 attenuation curve,
\begin{equation}
A_{\rm UV} = 12.56 \times E (B-V) = 2.26 \times A_{\rm H\alpha}
,\end{equation}
($A_{\rm UV} = 10.33 ~E (B-V)= 1.86 ~A_{\rm H\alpha}$ if the C00 law is used).\\

The corresponding value,  $A_{\rm UV}=2.7\pm0.4$ mag (2.23$\pm0.4$ for C00),  is plotted in Fig.\ref{Afuv-z} and is marginally  consistent with the other measures  performed in a UV selection or globally with the luminosity functions (and fully consistent if the C00 attenuation law is used).  \cite{kashino13} found    a larger value for the ratio $E(B-V)/E_g(B-V)$ reaching 0.7-0.8. Using this value for the color excess would lead to an attenuation of 4.6 mag, close to the values found for our IR selected samples, we do not favor this a high value as discussed below.\\

\subsection{Comparing dust attenuation in IR and UV or H$\alpha$ selected samples}
 \begin{figure}
   \centering
    \includegraphics[width=\columnwidth]{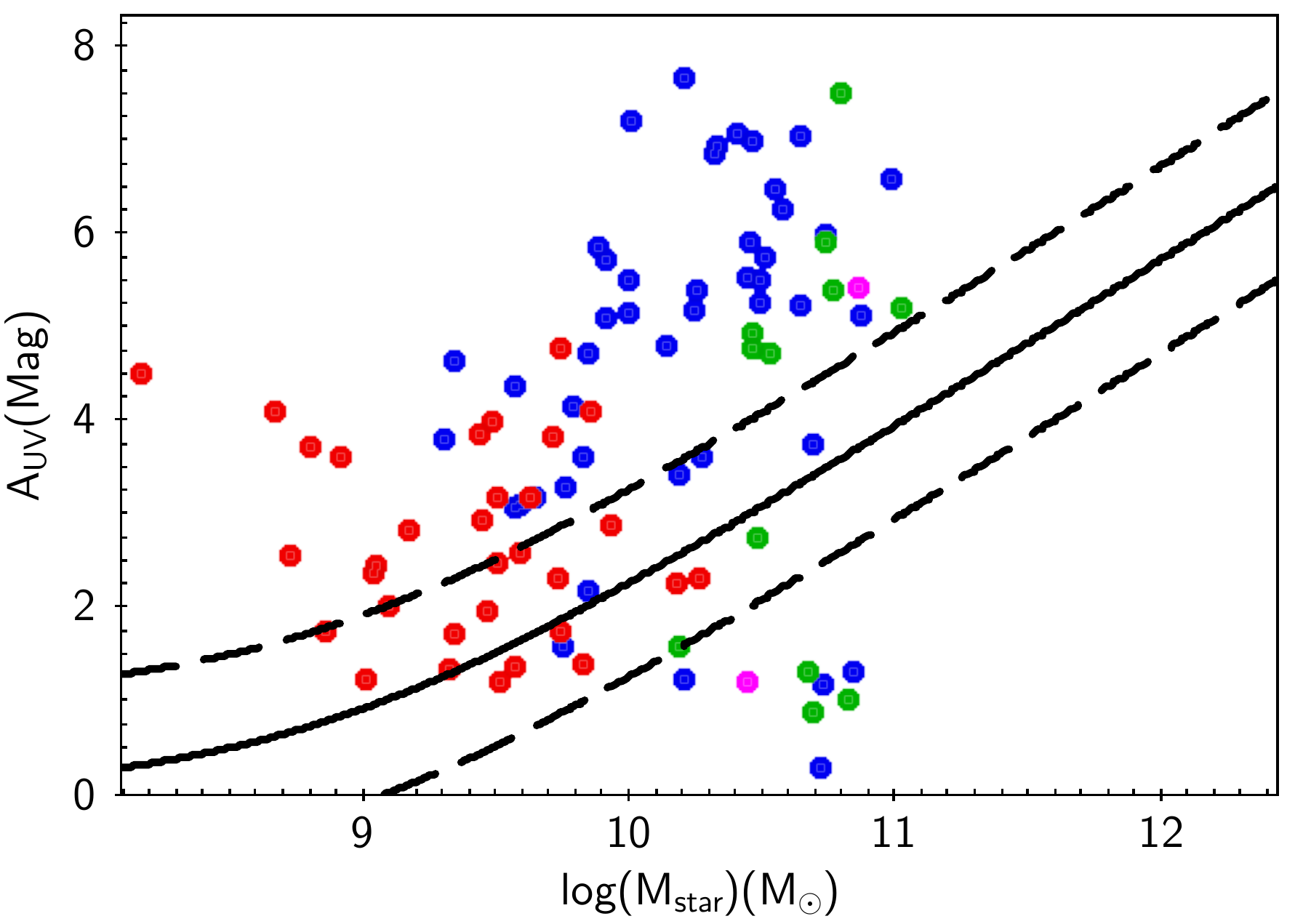}
    \caption{$A_{\rm UV}$ versus  $M_{\rm star}$ for starburst galaxies defined with a sSFR 0.6 dex higher than the average value found in each redshift bin. The color coding is the same as in Fig.\ref{Ldust-z}. The $A_{\rm UV}$-$M_{\rm star}$ relation assumed  by \citet{bernhard14} is plotted with a solid black line, and the 1$\sigma$ dispersion with dotted lines.}
    \label{auv-mstar-sb}
    \end{figure}

We have to  explain why the attenuation found in  IR selections is much higher than those found in UV or emission lines selected samples and on average in the universe.  With our  IR selection, we select galaxies dominating the star formation at the redshifts   considered. We find these galaxies experience a dust attenuation that is much higher than the average attenuation  in the universe or that is found in UV selected samples: the galaxies dominating the  star formation are not representative of the average attenuation   in the universe.  \\
Dust attenuation is found to  depend on the stellar mass  without a strong evolution with redshift  (see references above). We can try to use the relation linking  these two quantities  to account for the stellar mass distributions when we compare dust attenuation from different selections.\\
 Let us consider  the measure of \citet{heinis13} at  $z\sim1.5$:  the average stellar mass for our IR selection in the redshift bin 3 is $10^{10.9} M_{\odot}$. Using the relation found by \cite{heinis14}   between dust attenuation and $M_{\rm star}$,  we find that the average attenuation increases from 1.8 to 3.8 mag  if we only consider UV selected galaxies with an average stellar mass of $10^{10.9} M_{\odot}$, leading to  a value consistent with the measures obtained in  IR samples at the same redshift. \\
The same exercise can be performed for the H$\alpha$ selected sample of \citet{ibar13} at the same redshift. Using their relation  between attenuation in the H$\alpha$ line  and $M_{\rm star}$ for  $M_{\rm star}= 10^{10.9} M_{\odot}$, we find $A_{\rm H\alpha}=1.6$ mag, which translates to 3.6 mag in UV   a result again consistent with what is found for our IR selected samples of similar mass. These new values, which account for the stellar mass distribution, are also reported in Fig.\ref{Afuv-z}. Using the calibration of  stellar and gas color excess  of \citet{kashino13}  gives an attenuation of 5.8 mag (and 4.7 mag for the C00 law), much higher than that found in our IR selection. The value found by \citet{kashino13} may apply to galaxies experiencing a phase of intense star formation,  but it  seems to  overestimate the UV attenuation in our sample of galaxies. However an intermediate value  between 0.44 and 0.7-0.8 would also be consistent with our results. \\
 We find that considering the relation between dust attenuation and stellar mass helps at  reconciling measures of dust attenuation  in UV, H$\alpha,$  and IR selections. The phenomenological model proposed by  \cite{bernhard14}  to explain the evolution of UV and IR luminosity functions with  redshift is a good framework  to go ahead in our  analysis. Starting from  stellar mass distributions at different redshifts, they assume a SFR-$M_{\rm star}$ relation (the so-called main sequence) and the  relation between attenuation and stellar mass found by \cite{heinis14} with a significant dispersion ($\sigma=1$ mag, see Fig.\ref{auv-lir}) to predict $L_{\rm IR}$ and $L_{\rm UV}$ for each object. They are  able to reproduce the average trends found in the UV selections of \cite{heinis14} and their model is  well suited to simulate an IR selection.   We  apply  the same selection to  their simulated sample  as the selection we performed in Fig.\ref{Ldust-z} and Table 3. The  average  attenuation we get is equal respectively to 2.89, 3.93, 3.81, and 4.05 mag  for the 4 bins defined in Table 3, with a dispersion on the order of 1 mag. It is   in good agreement with our measurements as shown in Fig.\ref{Afuv-z}.\\
A major assumption of the model of \cite{bernhard14}, which we can try to check, is their recipe for dust attenuation. In their model they distinguish between galaxies on the main sequence and starbursting objects. The galaxies on the main sequence  follow the $A_{\rm UV}-M_{\rm star}$ relation described above and plotted in our Fig.\ref{auv-lir}. They are defined as starbursts sources   with a specific SFR (sSFR  defined as SFR/ $M_{\rm star}$) 0.6 dex above the average value of the main sequence. The attenuation of these starbursting objects is assumed to be  higher than that of galaxies on the main sequence and  the fraction of star formation attributed to the starburst is  completely obscured in their model.   We can check   if starbursting galaxies  from our sample exhibit a larger attenuation than the other galaxies. We adopt a similar  definition  to define starbursts  in each redshift bin by selecting sources with a sSFR 0.6 dex above  the average value  of the sSFR found in the bin. Our aim is is not to fully study the relation between SFR and $M_{\rm star}$ in our sample, and moreover our 8$\mu$m selection is likely to miss starbursting systems with an IR  luminosity close to the detection limit since these objects exhibit a lower 8$\mu$m rest frame to total IR luminosity \citep{murata14}. The stellar mass and dust attenuation of these galaxies with a large sSFR are reported in Fig.\ref{auv-mstar-sb}. We find that the starburst  galaxies are preferentially located above the average relation, thus exhibiting a large attenuation,  but  the effect is not systematic. These galaxies are neither numerous  nor attenuated enough   to explain the  'excess' of $\sim20\%$ of galaxies found in Sect. 4.3.3 to show an attenuation larger than that expected from  the average $A_{\rm UV}-M_{\rm star}$ relation: in our sample, galaxies with a more moderate star formation activity can   also experience  a large  dust attenuation.

\section{Conclusions}
We have studied the evolution of the amount of dust attenuation in a sample of 4077 galaxies selected in their rest-frame 8$\mu$m in the $AKARI$-NEP Deep Field observed with the IRC instrument. To complement  optical and near-infrared data, $GALEX$ and $Herschel$/PACS data are also added to the set of data.\\
Dust attenuation is measured from the fit of the UV-to-IR SED of each galaxy  with the CIGALE code. An AGN component is introduced from  the \cite{fritz06} library.  We  use different attenuation laws. The method of SED fitting  is optimized on a subsample of 106 galaxies with spectroscopic redshift. The choice of the attenuation law is found to affect only marginally the measure of  stellar mass and of  SFR, the latter being  essentially measured with the IR emission. Conversely, the amount of dust attenuation in the UV  rest frame at 150 nms varies with the choice of the  attenuation curve: adopting  the \cite{buat12}  law as a reference, a higher attenuation is found  with a SMC extinction law and a lower attenuation with the \cite{calzetti00} recipe, and inverse trends are found for the attenuation in the rest-frame V-band.  We find that the AGN contribution to the total IR emission is  low, but the introduction of AGN type 1 or type 2 templates in the fitting process affects the estimation of  dust attenuation. The  dust attenuation  is systematically higher with an AGN type 1 contribution. \\
Dust attenuation is found to increase with redshift  up to $z=1$ and the measurements are consistent with a constant attenuation in the redshift range 1-1.5, in agreement with previous measurements on a smaller range of  redshift. Dust attenuation  increases with both $L_{\rm IR}$ and $M_{\rm star}$. No trend with redshift is found when $A_{\rm UV}$ is plotted against $M_{\rm star}$. For a given $L_{\rm IR}$, we find the  dust attenuation  decreases slightly  when redshift increases in consistency with the large dispersion found when the attenuation is plotted against $L_{\rm IR}$. We find that the dust  attenuation is  larger than that measured in UV selected samples or samples selected in the H$\alpha$ line  at similar redshift, and that the dust attenuation is larger than the global attenuation in the universe measured with the IR and UV luminosity densities, the difference reaching $\sim 2 $ mag.  This large difference is well explained when considering the average  relation between dust attenuation and stellar mass: an IR selection is biased toward massive galaxies whereas UV selected galaxies exhibit a large range of stellar mass. Starbursting galaxies do not systematically exhibit a large attenuation. The phenomenological  model of \cite{bernhard14} is able to very well reproduce  the average measure of dust attenuation at different redshifts, but we find that the  relation between  $M_{\rm star}$ and $A_{\rm UV}$  more dispersed in our sample than is assumed in their model.

\begin{acknowledgements}
We thank Matthieu B\'ethermin for fruitful discussions and to have given us the opportunity to work with the simulated sample of \cite{bernhard14}. We also thank  Kazumi Murata for his help during  the definition of the samples. We are grateful to Jacopo Fritz for his useful comments about our treatment of the AGN contribution and to M\'ed\'eric Boquien for his help in running CIGALE. V. Buat and D. Burgarella acknowledge the support of the Institute of Space and Astronautical Science. V. Buat is also supported by  the Institut Universitaire de France. Kasia Malek  has been supported by the National Science Centre (grant UMO-2013/09/D/ST9/04030), and by the JSPS Strategic Young Researcher Overseas Visits Program for Accelerating Brain Circulation. Youichi Ohyama is supported by a MOST grant (100-2112-M-001-001-MY3).
      \end{acknowledgements}

\end{document}